\documentclass[12pt]{article}
\usepackage{amsfonts}
\newcommand{\eqdef}{\stackrel{\rm def}{=}}
\newcommand{\sfrac}[2]{{\textstyle \frac{#1}{#2}}}
\newcommand{\n}{\nonumber \\}
\newcommand{\dr}{\sqrt{\delta}}
\newcommand{\schoose}[2]{{\textstyle{#1\choose#2}}}
\newcommand{\bchoose}[2]{{#1\choose#2}}

\newtheorem{thm}{Theorem}[section]
\newtheorem{prop}[thm]{Proposition}

\newcommand{\maprightu}[2]
  {\smash{\mathop{\hbox to #1{\rightarrowfill}}\limits^{#2}}}
\newcommand{\maprightd}[2]
  {\smash{\mathop{\hbox to #1{\rightarrowfill}}\limits_{#2}}}

\newcommand{\mapdownr}[1]
  {\Bigg\downarrow\rlap{$\vcenter{\hbox{$\scriptstyle#1$}}$}}

\renewcommand{\theequation}{\arabic{section}.\arabic{equation}}

\begin{document}

\baselineskip=20pt

%%%%%%%%%%%%%%%%%%%%%%%%%%%%%%%%%%%%%%%%%%%%%%%%%%%%%%%%%%%%
%                                                          %
%  Title page                                              %
%                                                          %
%%%%%%%%%%%%%%%%%%%%%%%%%%%%%%%%%%%%%%%%%%%%%%%%%%%%%%%%%%%%
\newfont{\elevenmib}{cmmib10 scaled\magstep1}
\newcommand{\preprint}{
   \begin{flushleft}
     \elevenmib Yukawa\, Institute\, Kyoto\\
   \end{flushleft}\vspace{-1.3cm}
   \begin{flushright}\normalsize  \sf
     DPSU-04-1\\
     YITP-04-36\\
     {\tt hep-th/0407155} \\ July 2004
   \end{flushright}}
\newcommand{\Title}[1]{{\baselineskip=26pt
   \begin{center} \Large \bf #1 \\ \ \\ \end{center}}}
\newcommand{\Author}{\begin{center}
   \large \bf S.~Odake${}^a$ and R.~Sasaki${}^b$ \end{center}}
\newcommand{\Address}{\begin{center}
     $^a$ Department of Physics, Shinshu University,\\
     Matsumoto 390-8621, Japan\\
     ${}^b$ Yukawa Institute for Theoretical Physics,\\
     Kyoto University, Kyoto 606-8502, Japan
   \end{center}}
\newcommand{\Accepted}[1]{\begin{center}
   {\large \sf #1}\\ \vspace{1mm}{\small \sf Accepted for Publication}
   \end{center}}

\preprint
\thispagestyle{empty}
\bigskip\bigskip\bigskip

\Title{Equilibria of `Discrete' Integrable Systems and
Deformation of Classical Orthogonal Polynomials}
\Author

\Address

\begin{abstract}
The Ruijsenaars-Schneider systems 
are `discrete' version of the Calogero-Moser (C-M) systems in the sense that
the momentum operator $p$ appears in the Hamiltonians as a polynomial
in $e^{\pm\beta' p}$ ($\beta'$ is a deformation parameter) instead of
an ordinary polynomial in $p$ in the hierarchies of C-M systems.
We determine the polynomials describing the equilibrium positions
of the rational and trigonometric Ruijsenaars-Schneider systems
based on classical root systems.
These are deformation of the classical orthogonal polynomials,
the Hermite, Laguerre and Jacobi polynomials which describe the equilibrium
positions of the corresponding Calogero and Sutherland systems.
The orthogonality of the original polynomials is inherited by the deformed
ones which satisfy three-term recurrence and certain functional equations.
The latter reduce to the celebrated second order differential equations
satisfied by the classical orthogonal polynomials.
\end{abstract}

%%%%%%%%%%%%%%%%%%%%%%%%%%%%%%%%%%%%%%%%%%%%%%%%%%%%%%%%%%%%%%%
%                                                             %
%  1. Introduction                                            %
%                                                             %
%%%%%%%%%%%%%%%%%%%%%%%%%%%%%%%%%%%%%%%%%%%%%%%%%%%%%%%%%%%%%%%
\section{Introduction}
\label{intro}
\setcounter{equation}{0}

Exactly solvable or quasi-exactly solvable multi-particle
quantum mechanical systems  have many remarkable properties.
By definition, the entire (or a part of the) spectrum and
the corresponding eigenfunctions are calculable by algebraic means.
The corresponding classical systems share also many `quantum'
features. For example, the frequencies of small oscillations near
the classical equilibrium are `quantized' together with the eigenvalues
of the associated Lax matrices at the equilibrium. These phenomena
have been explored extensively for multi-particle dynamics based on root
systems, in particular, for the Calogero and Sutherland systems
\cite{Cal,Sut, CalMo} by Corrigan-Sasaki \cite{cs}.
Similar phenomena are also reported by Ragnisco-Sasaki \cite{rs} for
Ruijsenaars-Schneider systems
\cite{Ruij-Sch,vanDiejen1,vanDiejen2,vanDiejen0},
which are {\em deformation\/} of C-M systems.

In this paper we will discuss one special aspect of the classical
equilibria of exactly solvable systems based on
classical root systems, the rational
and trigonometric Ruijsenaars-Schneider
systems. Namely, the determination of the equilibrium positions and
their description in terms of certain polynomials.
It is known that for the Calogero and Sutherland systems,
the equilibrium positions are described by the zeros of the
classical orthogonal polynomials, {\em ie\/} the
Hermite, Laguerre, Chebyshev, Legendre, Gegenbauer
and Jacobi polynomials
\cite{cs,calmat,os}.
The Ruijsenaars-Schneider systems are `good' deformation of the
Calogero and Sutherland systems.
Here is one interesting evidence.
It was known \cite{AMOS} that the singular vectors of the Virasoro and $W_N$
algebras, in the free field representation, are related to Jack polynomials
\cite{Jack}, the quantum eigenfunctions of the $A$ type Sutherland
systems. The deformed Virasoro and
$W_N$ algebras were discovered by using the relation between the
Sutherland system and the trigonometric Ruijsenaars-Schneider system
of the $A$ type root system \cite{AKOS}.
Therefore it is expected that equilibrium positions of the rational and
trigonometric Ruijsenaars-Schneider systems would give certain `good'
deformation of the classical orthogonal polynomials:
\medskip
\begin{eqnarray*}
  \framebox{\shortstack{$\,$Calogero-Sutherland$\,$ \\ \\ systems}}
  &\maprightu{45mm}{\mbox{equilibrium positions}}&
  \framebox{\shortstack{classical \\ \\
      orthogonal polynomials}}\\
  \mapdownr{\mbox{`good' deformation}}\hspace{20mm}&&
%  \hspace{15mm}\mapdownr{\mbox{`good' deformation expected}}\\
  \hspace{15mm}\mapdownr{\makebox{
     \shortstack{`good' deformation \\ expected}}}\\
  \hspace*{-5mm}
  \framebox{\shortstack{Ruijsenaars-Schneider \\ \\ systems}}
  &\maprightu{45mm}{\mbox{equilibrium positions}}&
%  \framebox{\shortstack{deformation of \\ \\ these polynomials}}
  \framebox{\shortstack{$\quad$ deformed classical \\ \\
           orthogonal polynomials}}
\end{eqnarray*}

\noindent
In Ragnisco-Sasaki paper \cite{rs}, based on numerical analysis,
the explicit forms of the lower degree members of the one-parameter
deformation of the Hermite and Laguerre polynomials were presented.
The present authors continued the numerical analysis and obtained
the explicit forms of the lower degree members of the one and/or two-parameter
deformation of the Hermite polynomial, one, two and/or three-parameter
deformation of the Laguerre polynomial, one-parameter deformation of
the Jacobi (and Gegenbauer and Legendre) polynomials.
They are also polynomials, or rational functions in the deformation
parameter(s) with {\em integer coefficients\/}.

Remarkably the orthogonality of the original polynomials is inherited by
the deformed ones.
The equations determining the equilibrium can be reformulated as
functional equations determining the polynomials.
These functional equations are  difference analogs of the celebrated second
order differential equations satisfied by the classical orthogonal
polynomials.
Three term recurrence for the deformed polynomials, the necessary and
sufficient condition for  orthogonality, can be derived from these
functional equations.
Dynamical stability of the Hamiltonian system, or bounded-from-belowness
of its potentials, is achieved by restricting the parameter space of the
coupling constants, usually by {\em positive coupling constants\/},
which in turn guarantees the positive definiteness of the inner product
governing the orthogonal polynomials, the deformed as well as undeformed.
These deformed polynomials are {\em not\/} the so-called $q$-deformed
versions of the above classical polynomials \cite{And-Ask-Roy}.

This paper is organized as follows.
In section 2, first we recall the essence of the Calogero-Sutherland
systems and  their equilibria, which are described by the Hermite,
Laguerre and Jacobi polynomials. Next the Hamiltonians and potentials of
the  Ruijsenaars-Schneider (R-S) systems are briefly recapitulated,
and two types of the rational systems and one trigonometric systems
for the classical root systems are introduced.
Then the equations for their equilibrium positions are written down.
For later use we review the relation between the orthogonal polynomials
and the three-term recurrence.
Section 3, 4 and 5 give the main results of this paper.
In section 3 we determine the equilibrium positions of the rational R-S
systems for the $A$ type root system and the deformation of the Hermite
polynomial is presented.
For one parameter deformation case, we derive the explicit forms of the
generating function and the weight function of the inner product.
In section 4 equilibrium positions of the rational R-S systems for
$B,C,D,BC$ type root system are determined and the deformation of the
Laguerre polynomial is presented together with the explicit forms of the
functional equations and three-term recurrence. In section 5 equilibrium
positions of the trigonometric R-S systems for the
$B,C,D,BC$ type root system are determined and the deformation of
the Jacobi (and Gegenbauer) polynomial is presented together with the
explicit forms of the functional equations and three-term recurrence.
Classical orthogonal polynomials, {\em eg\/}, the Hermite and Laguerre, 
satisfy many interesting identities among themselves. Those having a root
theoretic explanation ({\em folding\/}) are shown to be preserved after 
deformation. 
Final section is devoted to summary and comments. Identification of
the deformed orthogonal polynomials within the so-called Askey-scheme 
of hypergeometric orthogonal polynomials \cite{koeswart} is reported
here. 
The relation between the functional equation and the three-term
recurrence is discussed in the Appendix.

%%%%%%%%%%%%%%%%%%%%%%%%%%%%%%%%%%%%%%%%%%%%%%%%%%%%%%%%%%%%%%%
%                                                             %
%  2. Potentials and Equilibrium Positions                    %
%                                                             %
%%%%%%%%%%%%%%%%%%%%%%%%%%%%%%%%%%%%%%%%%%%%%%%%%%%%%%%%%%%%%%%
\section{Potentials and Equilibrium Positions}
\setcounter{equation}{0}

In this section we set up models and present the equations for their
equilibrium positions.
We consider a multi-particle classical mechanics governed by a classical
Hamiltonian $H(p,q)$.
The dynamical variables are the coordinates
\(\{q_{j}|\,j=1,\ldots,r\}\) and their canonically conjugate momenta
\(\{p_{j}|\,j=1,\ldots,r\}\).
These will be  denoted by vectors in \(\mathbb{R}^{r}\)
\[
   q={}^t(q_{1},\cdots,q_{r}),\quad p={}^t(p_{1},\cdots,p_{r}),
\]
in which $r$ is the number of particles (and it is also the rank of the
underlying root system $\Delta$ except for the $A$ case).
The canonical equations of motion are
\begin{equation}
  \dot{q}_j={\partial H(p,q)\over{\partial p_j}}\,,\quad
  \dot{p}_j=-{\partial H(p,q)\over{\partial q_j}}\,.
\end{equation}
The equilibrium positions are the stationary solution
\begin{equation}
  p=0\,,\quad q=\bar{q}\,,
  \label{stasol}
\end{equation}
in which $\bar{q}$ satisfies
\begin{equation}
  {\partial H(0,q)\over{\partial q_j}}\Biggm|_{q=\bar{q}}=0\,,\quad
  (j=1,\ldots,r).
  \label{staeq}
\end{equation}

We will discuss Ruijsenaars-type models associated with the
{\em classical\/} root systems, namely the $A_{r-1}$, $B_r$, $C_r$,
$D_r$ and $BC_r$.
The fact that all the roots of the classical root systems are neatly
expressed in terms of the orthonormal basis of $\mathbb{R}^r$ makes
formulation much simpler than those systems based on the exceptional
root systems.
The sets of positive roots of the classical root systems are
\begin{eqnarray*}
  A_{r-1}&:&\Delta_+=\{e_j-e_k|1\leq j<k\leq r\},\\
  B_r&:&\Delta_{L+}=\{e_j\pm e_k|1\leq j<k\leq r\},\quad
  \Delta_{S+}=\{e_j|1\leq j\leq r\},\\
  C_r&:&\Delta_{S+}=\{e_j\pm e_k|1\leq j<k\leq r\},\quad
  \Delta_{L+}=\{2e_j|1\leq j\leq r\},\\
  D_r&:&\Delta_+=\{e_j\pm e_k|1\leq j<k\leq r\},\\
  BC_r&:&\Delta_{M+}=\{e_j\pm e_k|1\leq j<k\leq r\},\\
  &&
  \Delta_{S+}=\{e_j|1\leq j\leq r\},\quad
  \Delta_{L+}=\{2e_j|1\leq j\leq r\},
\end{eqnarray*}
where $\{e_j\}$ is an orthonormal basis of $\mathbb{R}^r$.
The subscripts $L,M$ and $S$ stand for long, middle and short roots,
respectively.

\bigskip
It is well known that the non-simply laced root systems are obtained from
simply laced ones by {\em folding\/}.
In the present case, the relevant ones are:
\begin{equation}
  A_{2r-1}\to C_r,\quad
  D_{r+1}\to B_r,\quad
  A_{2r}\to BC_r.
  \label{foldtypes}
\end{equation}
At the level of the dynamical variables and Hamiltonians, the above foldings
are realised as:
\begin{eqnarray}
  A_{2r-1}\to C_r &:& p_{2r+1-j}= -p_j, \quad q_{2r+1-j}= -q_j,\quad
  (j=1,\ldots,r),
  \label{acfold}\\
  D_{r+1}\to B_r &:& p_{r+1}=q_{r+1}=0,
  \label{dbfold}\\
  A_{2r}\to BC_r &:& p_{2r+2-j}= -p_j, \quad q_{2r+2-j}= -q_j,\ \
  (j=1,\ldots,r),\ \ p_{r+1}=q_{r+1}=0.
  \label{bcfold}
\end{eqnarray}

%%%%%%%%%%%%%%%%%%%%%%%%%%%%
%  2.1                     %
%%%%%%%%%%%%%%%%%%%%%%%%%%%%
\subsection{Calogero and Sutherland Systems}
\label{cssub}

For later comparison, we summarise the Calogero and Sutherland systems
associated with the {\em classical\/} root systems only, 
namely the $A_{r-1}$, $B_r$, $C_r$, $D_r$ and $BC_r$.

The Hamiltonian of the classical Calogero and Sutherland systems is
\begin{equation}
  H_{\rm CS}(p,q)={1\over2}\sum_{j=1}^r p_j^2+V_{\rm C}(q),
  \label{HCS}
\end{equation}
where the classical potential $V_{\rm C}$ is given below explicitly.
For all cases this classical potential $V_{\rm C}$ can be written
in terms of the {\em prepotential\/}  $W(q)$ \cite{bms}
\begin{equation}
  V_{\rm C}(q)
  =\frac12\sum_{j=1}^r\left(\frac{\partial W(q)}{\partial q_j}\right)^{\!2}.
  \label{V=W'^2}
\end{equation}
The equations for the equilibrium positions (\ref{staeq}) reduce to
the following equations:
\begin{equation}
  \frac{\partial W(q)}{\partial q_j}\Biggm|_{q=\bar{q}}=0,\quad(j=1,\ldots,r).
  \label{W'=0}
\end{equation}

%%%%%%%%%%%%%%%%%%%%%%%%%%%%
%  2.1.1                   %
%%%%%%%%%%%%%%%%%%%%%%%%%%%%
\subsubsection{Calogero Systems}

The classical potential $V_{\rm C}$ and prepotential $W$  are
\begin{eqnarray}
  V_{\rm C}(q)&\!\!=\!\!&\frac{\omega^2}{2}\sum_{j=1}^rq_j^2
  +\frac12\sum_{\rho\in\Delta_+}\frac{g_{\rho}^2\rho^2}{(\rho\cdot q)^2}\,,\\
  W(q)&\!\!=\!\!&-\frac{\omega}{2}\sum_{j=1}^rq_j^2
  +\sum_{\rho\in\Delta_+}g_{\rho}\log|(\rho\cdot q)|\,,
\end{eqnarray}
where $\omega$ is the (positive) frequency of the harmonic confining potential
and $g_{\rho}$ are real positive coupling constants depending
on the length of the roots.
They are: one coupling  $g$ for all roots for the $A_{r-1}$ and $D_r$,
two independent couplings
$g_L$ and $g_S$ for $B_r$ and $C_r$ corresponding to the long and short
roots, respectively.
These conventions are the same for all other types
of potentials considered in this paper.
For example, the $C_r$ model is
\begin{eqnarray*}
  C_r&:&V_{\rm C}(q)=\frac{\omega^2}{2}\sum_{j=1}^rq_j^2
  +\frac{g_S^{\,2}}{2}\sum_{j,k=1\atop j\neq k}^r
  \left({1\over{(q_j-q_k)^2}}+{1\over{(q_j+q_k)^2}}\right)
  +\frac{g_L^{\,2}}{2}\sum_{j=1}^r{1\over{q_j^2}}\,,\\
  &&W(q)=-\frac{\omega}{2}\sum_{j=1}^rq_j^2
  +\frac{g_S}{2}\sum_{j,k=1\atop j\neq k}^r\log|q_j^2-q_k^2|
  +g_L\sum_{j=1}^r\log|2q_j|\,.
\end{eqnarray*}
There is no distinction between the rational $B_r$ and $C_r$ models
because of the replacement $g_S\leftrightarrow g_L$.
The $D_r$ model can be considered as a special case of the $B_r$
with $g_L=g$ and $g_S=0$.

The systems obtained by {\em folding\/} (\ref{acfold})--(\ref{bcfold})
have a special ratio of couplings. They are
\begin{equation}
  \begin{array}{lcllcl}
  \mbox{folded}\ C_r&:& (g_L,g_S)=({1\over2},1)g, &
  \mbox{folded}\ B_r&:& (g_L,g_S)=(1,2)g,\\[3pt]
  \mbox{folded}\ BC_r&:& (g_L,g_M,g_S)=({1\over2},1,1)g.&&
  \end{array}
  \label{foldrel1}
\end{equation}

The equations for the equilibrium position (\ref{W'=0}) are
\begin{eqnarray}
  A_{r-1}&:&\sum_{k=1\atop k\ne j}^r{1\over{\bar{q}_j-\bar{q}_k}}
  =\frac{\omega}{g}\,\bar{q}_j\,,
  \label{herdet}\\
  B_r&:&\sum_{k=1\atop k\ne j}^r{2\bar{q}_j\over{\bar{q}_j^2-\bar{q}_k^2}}
  =\frac{\omega}{g_L}\,\bar{q}_j
  -\frac{g_S}{g_L}\,{1\over{\bar{q}_j}}\,,
  \label{lagdetB}\\
  C_r&:&\sum_{k=1\atop k\ne j}^r{2\bar{q}_j\over{\bar{q}_j^2-\bar{q}_k^2}}
  =\frac{\omega}{g_S}\,\bar{q}_j
  -\frac{g_L}{g_S}\,{1\over{\bar{q}_j}}\,,
  \label{lagdetC}\\
  D_r&:&\sum_{k=1\atop k\ne j}^r{2\bar{q}_j\over{\bar{q}_j^2-\bar{q}_k^2}}
  =\frac{\omega}{g}\,\bar{q}_j\,.
  \label{lagdetD}
\end{eqnarray}
They determine the zeros of the Hermite and Laguerre polynomials.
In other words, if we define \(\bar{q}_j=\sqrt{\frac{g}{\omega}}\,y_j\)
for $A_{r-1}$, then the polynomial having $\{y_j\}$ as zeros is the
Hermite polynomial \cite{calmat,szego,cs}:
\begin{equation}
  2^r\prod_{j=1}^r(x-y_j)
  =r!\sum_{j=0}^{[\frac{r}{2}]}\frac{(-1)^j(2x)^{r-2j}}{j!(r-2j)!}
  \eqdef H_r(x).
  \label{hermdef}
\end{equation}
For the $C_r$ (or $B_r$) model let us define
\(\bar{q}_j=\sqrt{\frac{g_S}{\omega}}\,y_j\),
\(\alpha=\frac{g_L}{g_S}-1\) then having $\{y_j^2\}$ as
zeros is the Laguerre polynomial \cite{calmat,szego,cs}:
\begin{equation}
  \frac{(-1)^r}{r!}\prod_{j=1}^r(x-y_j^2)=\sum_{j=0}^r
  \bchoose{r+\alpha}{r-j}\frac{(-x)^j}{r!}\eqdef L^{(\alpha)}_r(x).
  \label{lagdef}
\end{equation}
For the $D_r$ root system, it is the Laguerre polynomial $L^{(-1)}_r(x)$.

The identities between the Hermite and Laguerre polynomials
\begin{eqnarray}
  2^{-2r}H_{2r}(x)&\!\!=\!\!&(-1)^r r!\,L_r^{(-\frac12)}(x^2),
  \label{LHiden}\\
  2^{-2r-1}H_{2r+1}(x)&\!\!=\!\!&x\,(-1)^r r!\,L_r^{(\frac12)}(x^2),
  \label{LHiden2}
\end{eqnarray}
are well-known. The former identity (\ref{LHiden}) for the even degree
Hermite polynomial can be understood as a consequence of the folding of
the root system $A_{2r-1}\to C_r$, see (\ref{acfold}).
Likewise
the latter identity (\ref{LHiden2}) for the odd degree Hermite
polynomial can be understood as a consequence of the folding of
the root system $A_{2r}\to BC_r$, see (\ref{bcfold}).
Next let us consider the folding $D_{r+1}\to B_r$ (\ref{dbfold}),
which leads to the identity \cite{cs} among the Laguerre polynomials of
different indices:
\begin{equation}
  (r+1)\,L^{(-1)}_{r+1}(x)=-x\,L^{(1)}_r(x).
\label{lagsiden}
\end{equation}
We will see that these identities (\ref{LHiden}), (\ref{LHiden2}) and
(\ref{lagsiden}) are also nicely {\em deformed\/} with one parameter 
(\ref{LHidendef1}), (\ref{LHiden2def1}) and (\ref{deforLHiden}) 
and with two parameters 
(\ref{LHidendef2}), (\ref{LHiden2def2}) and (\ref{2deforLHiden}).

%%%%%%%%%%%%%%%%%%%%%%%%%%%%
%  2.1.2                   %
%%%%%%%%%%%%%%%%%%%%%%%%%%%%
\subsubsection{Sutherland Systems}

The classical potential $V_{\rm C}$ and prepotential $W$ are
(except for $V_{\rm C}$ of $BC_r$)
\begin{eqnarray}
  V_{\rm C}(q)&\!\!=\!\!&\frac12\sum_{\rho\in\Delta_+}
  \frac{g_{\rho}^2\rho^2}{\sin^2(\rho\cdot q)}\,,\\
  W(q)&\!\!=\!\!&
  \sum_{\rho\in\Delta_+}g_{\rho}\log|\sin(\rho\cdot q)|\,,
\end{eqnarray}
where $g_{\rho}$ are real positive coupling constants.
The classical potential $V_{\rm C}$ of the $BC_r$ model is given by
\begin{eqnarray}
  BC_r&:&V_{\rm C}(q)=
  \frac{g_M^{\,2}}{2}\sum_{j,k=1\atop j\neq k}^r\left(
  \frac{1}{\sin^2(q_j-q_k)}+\frac{1}{\sin^2(q_j+q_k)}\right)\nonumber\\
  &&\phantom{V_{\rm C}(q)=}\quad
  +2g_L^{\,2}\sum_{j=1}^r\frac{1}{\sin^22q_j}
  +\frac{g_S(g_S+2g_L)}{2}\sum_{j=1}^r\frac{1}{\sin^2q_j}\,.
\end{eqnarray}
The $B_r$ ($C_r$) potential is obtained by setting $g_L=0$, $g_M\to g_L$,
($g_S=0$, $g_M\to g_S$).

The equations for the equilibrium position (\ref{W'=0}) are
\begin{eqnarray}
  A_{r-1}&:&
  \sum_{k=1\atop k\neq j}^r\cot(\bar{q}_j-\bar{q}_k)=0,
  \label{chedet}\\
  B_r&:&
  \sum_{k=1\atop k\neq j}^r
  \Bigl(\cot(\bar{q}_j-\bar{q}_k)+\cot(\bar{q}_j+\bar{q}_k)\Bigr)
  =-\frac{g_S}{g_L}\cot \bar{q}_j,
  \label{jacobidefB}\\
  C_r&:&
  \sum_{k=1\atop k\neq j}^r
  \Bigl(\cot(\bar{q}_j-\bar{q}_k)+\cot(\bar{q}_j+\bar{q}_k)\Bigr)
  =-2\frac{g_L}{g_S}\cot 2\bar{q}_j,
  \label{jacobidefC}\\
  BC_r&:&
  \sum_{k=1\atop k\neq j}^r
  \Bigl(\cot(\bar{q}_j-\bar{q}_k)+\cot(\bar{q}_j+\bar{q}_k)\Bigr)
  =-\frac{g_S}{g_M}\cot \bar{q}_j-2\frac{g_L}{g_M}\cot 2\bar{q}_j,
  \label{jacobidef}\\
  D_r&:&
  \sum_{k=1\atop k\neq j}^r
  \Bigl(\cot(\bar{q}_j-\bar{q}_k)+\cot(\bar{q}_j+\bar{q}_k)\Bigr)
  =0.
  \label{jacobidefD}
\end{eqnarray}
The equilibrium positions of the $A_{r-1}$ model are related to the Chebyshev
polynomial and those of the other models are related to the Jacobi
polynomials.

For the $A_{r-1}$, the equilibrium positions are
``{\em equally-spaced\/}" and translational invariant,
\begin{equation}
  \bar{q}=\frac{\pi}{r}\ {}^t(r,r-1,\cdots,1)+\xi\ {}^t(1,1,\cdots,1),\qquad
  \xi\in\mathbb{R}:\mbox{arbitrary}.
  \label{eqspaced1}
\end{equation}
We choose this constant shift $\xi$ such that the ``center of mass"
coordinate vanishes, $\displaystyle \sum_{j=1}^r\bar{q}_j=0$:
\begin{equation}
  \bar{q}_j=\frac{\pi(r+1-j)}{r}-\frac{\pi(r+1)}{2r}=
  \frac{\pi}{2}-\frac{\pi(2j-1)}{2r}=-\bar{q}_{r+1-j},
  \quad(j=1,\ldots,r).
  \label{eqspaced}
\end{equation}
Then the degree $r$ (the dimension of the vector
representation) polynomial in $x$, having zeros at $\{\sin\bar{q}_j\}$,
\begin{equation}
  2^{r-1}\prod_{j=1}^r(x-\sin\bar{q}_j)
  =2^{r-1}\prod_{j=1}^r\Bigl(x-\cos{\pi(2j-1)\over{2r}}\Bigr)
  \eqdef T_r(x)\,,
  \label{archev}
\end{equation}
is the Chebyshev polynomial of the first kind,
$T_n(\cos\varphi)=\cos(n\varphi)$.

For the solution $\{\bar{q}_j\}$ of  the $BC_r$ (\ref{jacobidef}),
$\cos 2\bar{q}_j$ is the zero  of the Jacobi polynomial
$P_r^{(\alpha,\beta)}(x)$
with $\alpha=\frac{g_S}{g_M}+\frac{g_L}{g_M}-1$ and
$\beta=\frac{g_L}{g_M}-1$,
\begin{equation}
  2^{-r}\schoose{\alpha+\beta+2r}{r}
  \prod_{j=1}^r(x-\cos 2\bar{q}_j)
  =\sum_{j=0}^r\schoose{\alpha+r}{r-j}
  \schoose{\alpha+\beta+r+j}{j}2^{-j}(x-1)^j
  \eqdef P_r^{(\alpha,\beta)}(x).
  \label{Pdef}
\end{equation}
It is easily shown that $\bar{q}'_j=\frac{\pi}{2}-\bar{q}_j$ satisfies
(\ref{jacobidef}) with $\alpha\leftrightarrow\beta$, which implies
$P_n^{(\alpha,\beta)}(-x)=(-1)^nP_n^{(\beta,\alpha)}(x)$.
For the solution $\{\bar{q}_j\}$ of the $C_r$ (\ref{jacobidefC}),
$\cos 2\bar{q}_j$ is the zero  of the Gegenbauer polynomial
$C_r^{(\alpha+\frac12)}(x)$
with $\alpha=\frac{g_L}{g_S}-1$,
\begin{equation}
  2^r\schoose{\alpha-\frac12+r}{r}
  \prod_{j=1}^r(x-\cos 2\bar{q}_j)
  =\schoose{2\alpha+r}{r}\schoose{\alpha+r}{r}^{-1}
  \sum_{j=0}^r\schoose{\alpha+r}{r-j}
  \schoose{\alpha+\beta+r+j}{j}2^{-j}(x-1)^j
  \eqdef C^{(\alpha+\frac12)}_r(x).
  \label{Gegendef}
\end{equation}
This is a special case of $P^{(\alpha,\beta)}_r(x)$ with another
normalisation,
\begin{equation}
  C^{(\alpha+\frac12)}_r(x)
  =\schoose{2\alpha+r}{r}\schoose{\alpha+r}{r}^{-1}
  P^{(\alpha,\alpha)}_r(x).
\end{equation}
For the solution $\{\bar{q}_j\}$  of the $B_r$  (\ref{jacobidefB}),
$\cos 2\bar{q}_j$ is the zero  of $P_r^{(\alpha,-1)}(x)$
with $\alpha=\frac{g_S}{g_L}-1$.
For the solution $\{\bar{q}_j\}$  of the $D_r$  (\ref{jacobidefD}),
$\cos 2\bar{q}_j$ is the zero  of $P_r^{(-1,-1)}(x)$.

The known identities between the Chebyshev and Jacobi polynomials and
between the Jacobi polynomials can be understood as consequences of the
{\em folding\/}:
\begin{eqnarray}
  A_{2r-1}\to C_r&:&\,\,\, 2^{1-2r}T_{2r}(x)=
  (-1)^r{\textstyle{2r-1\choose r}}^{-1}
  P_r^{(-\frac12,-\frac12)}(1-2x^2),
  \label{chebjaciden1}\\
  A_{2r}\to BC_r&:&2^{-2r}T_{2r+1}(x)=
  x\,(-1)^r{\textstyle{2r\choose r}}^{-1}
  P_r^{(\frac12,-\frac12)}(1-2x^2),
  \label{chebjaciden2}\\
  D_{r+1}\to B_r&:&{2(r+1)}P_{r+1}^{(-1,-1)}(x)=
  r(x-1)P_r^{(1,-1)}(x).
  \label{gegiden}
\end{eqnarray}
We will see in the following that the first two identities are not deformed but
the third one is nicely deformed (\ref{Jacidendef}).

The Gegenbauer polynomial and the Jacobi polynomial are also related by the
quadratic transformations :
\begin{eqnarray}
  C_{2n}^{(\alpha+\frac12)}(x)&\!\!=\!\!&
  2^{2n}\schoose{\alpha-\frac12+n}{n}\schoose{2n}{n}^{-1}
  P_n^{(\alpha,-\frac12)}(2x^2-1)\,,
  \label{genjaciden1}\\
  C_{2n+1}^{(\alpha+\frac12)}(x)&\!\!=\!\!&
  2^{2n+1}\schoose{\alpha+\frac12+n}{n+1}\schoose{2n+1}{n}^{-1}
  xP_n^{(\alpha,\frac12)}(2x^2-1)\,.
  \label{genjaciden2}
\end{eqnarray}
However these identities do not seem to have a folding type explanation.
Indeed the deformed Gegenbauer polynomials
$C_n^{(\alpha+\frac12)}(x,\delta)$ (\ref{gegendef}),
$\tilde{C}_n^{(\alpha+\frac12)}(x,\delta)$ (\ref{gegentdef}) and the
deformed Jacobi polynomial $P_n^{(\alpha,\beta)}(x,\delta)$
(\ref{defjacobidef}) do not satisfy this type of identities for generic
$\delta$.

%%%%%%%%%%%%%%%%%%%%%%%%%%%%
%  2.2                     %
%%%%%%%%%%%%%%%%%%%%%%%%%%%%
\subsection{Ruijsenaars-type Systems}

Following Ruijsenaars-Schneider \cite{Ruij-Sch} and
van Diejen \cite{vanDiejen1}, the Hamiltonian of the Ruijsenaars systems is
(the deformation parameter $\beta'$ of $e^{\pm\beta' p}$ is set to unity,
$\beta'=1$)
\begin{equation}
  H(p,q)=\sum_{j=1}^r\left( \cosh p_j\,\sqrt{V_j(q)\,{V}_j^*(q)}
  -\frac12\Bigl(V_j(q)+{V}_j^*(q)\Bigr)\right).
  \label{Ham}
\end{equation}
The form of the function $V_j=V_j(q)$ and its complex conjugate
${V}_j^*$ are determined by the root system $\Delta$ as :
\begin{eqnarray}
  A_{r-1} &:& V_j(q)=w(q_j)\prod_{k=1\atop k\neq j}^r v(q_j-q_k)\,,
  \qquad(j=1,\ldots,r),
  \label{AVform}\\
  B_r,\,C_r,\,D_r,\,BC_r &:&
  V_j(q)=w(q_j)\prod_{k=1\atop k\neq j}^r v(q_j-q_k)\,v(q_j+q_k)\,,
  \qquad(j=1,\ldots,r).
  \label{otherVform}
\end{eqnarray}
The elementary potential functions $v$ and $w$ depend on the nature of
interactions (rational, trigonometric, {\em etc\/}) and the root
system
$\Delta$. Their explicit forms will be given below.
When $V$ satisfies the simple identity
\(
  \sum_{j}\left(V_j(q)+V^*_j(q)\right)=\mbox{const.}\,,
\)
the Hamiltonian (\ref{Ham}) could be replaced by a simpler one
\begin{equation}
  H'(p,q)=\sum_{j=1}^r\cosh p_j\,\sqrt{V_j(q)\,{V}_j^*(q)}\,,
 \label{Ham2}
\end{equation}
which is obviously positive definite and
usually used as a starting point for the trigonometric
(hyperbolic) interaction theory.

The above Hamiltonian (\ref{Ham}) is a hyperbolic function of the
momentum operator $p$ instead of a polynomial in the hierarchy of C-M
systems or other ordinary dynamical systems.
In quantum theoretical setting this Hamiltonian causes finite shifts of
the wavefunction in the imaginary direction, {\em ie\/}
$\cosh p\, \psi(q)$$=\frac12(\psi(q- i\hbar)+\psi(q+ i\hbar))$,
in which $\hbar$ is Planck's constant. This is why the R-S systems are
sometimes called `discrete' dynamical systems.

The equation (\ref{staeq}) of equilibrium positions (\ref{stasol}) can
be simplified in the following way.
By expanding the Hamiltonian around the stationary solution
(\ref{stasol}), we obtain
\begin{equation}
  H(p,q)=K(p)+P(q)+\mbox{higher order terms in } p,
  \quad  K(p)=\frac12\sum_{j=1}^r|\,V_j(\bar{q})\,|\ p_j^2\,,
\end{equation}
and the `potential' $P$ is given by
\begin{equation}
  P(q)=\sum_{j=1}^r
  \left(\sqrt{V_j(q)\,{V}_j^*(q)}
  -\frac12\Bigl(V_j(q)+{V}_j^*(q)\Bigr)\right)
  =-\frac12\sum_{j=1}^r\left(\sqrt{V_j(q)}-\sqrt{V_j^*(q)}\right)^{\!2}.
  \label{potP}
\end{equation}
This should be compared with the classical potential in
the Calogero-Sutherland systems (\ref{V=W'^2}).
It is obvious that the equilibrium is achieved at the point(s) in which
all the functions $V_j$ become {\em real\/} and {\em positive\/}:
\begin{equation}
  V_j(\bar{q})=V_j^*(\bar{q})>0\,,\qquad(j=1,2,\ldots,r).
  \label{equeq}
\end{equation}
The equilibrium point is the absolute minimum of the potential $P$.
The system of equations (\ref{equeq}) is invariant under any permutation
of $\{\bar{q}_j\}$. For $v$ and $w$ considered in this paper, they are also
invariant under the transformation $q\rightarrow q'=-q$.
Except for the $A$ case, they are also invariant under the
transformation
$q\rightarrow q'={}^t(\epsilon_1q_1,\cdots,\epsilon_rq_r)$
with $\epsilon_i=\pm 1$.

The functions $v$ and $w$ considered in this paper have
properties
\begin{equation}
  v(-x)=v^*(x),\quad w(-x)=w^*(x),
\end{equation}
which allow the {\em folding\/}
(\ref{acfold})--(\ref{bcfold}) of the Ruijsenaars type Hamiltonians:
\begin{eqnarray}
  H^{A_{2r-1}}(p,q)\Bigm|_{p_{2r+1-j}=-p_j\atop q_{2r+1-j}=-q_j}
  =2\tilde{H}(p,q),&&
  \tilde{v}(x)=v^{A}(x),\quad\tilde{w}(x)=w^A(x)v^A(2x),
  \label{AtoC}\\
  H^{A_{2r}}(p,q)\Bigm|_{p_{2r+2-j}=-p_j\atop q_{2r+2-j}=-q_j}
  =2\tilde{H}(p,q),&&
  \tilde{v}(x)=v^{A}(x),\quad\tilde{w}(x)=w^A(x)v^A(x)v^A(2x),\quad
  \label{AtoBC}\\
  H^{D_{r+1}}(p,q)\Bigm|_{p_{r+1}=0,q_{r+1}=0}
  =\tilde{H}(p,q),&&
  \tilde{v}(x)=v^{D}(x),\quad\tilde{w}(x)=w^D(x)v^D(x)^2,
  \label{DtoB}
\end{eqnarray}
where $\tilde{H}$ is
\begin{eqnarray}
  \tilde{H}(p,q)&\!\!=\!\!&\sum_{j=1}^r\left( \cosh p_j\,
  \sqrt{\tilde{V}_j(q)\,\tilde{V}_j^*(q)}
  -\frac12\Bigl(\tilde{V}_j(q)+\tilde{V}_j^*(q)\Bigr)\right),\nonumber\\
  \tilde{V}_j(q)&\!\!=\!\!&\tilde{w}(q_j)\prod_{k=1\atop k\neq j}^r
  \tilde{v}(q_j-q_k)\,\tilde{v}(q_j+q_k)\,.
\end{eqnarray}
The folded systems (\ref{AtoC}), (\ref{AtoBC}) and
(\ref{DtoB}) correspond to the folding $A_{2r-1}\to C_r$, 
$A_{2r}\to BC_r$ and $D_{r+1}\to B_r$, respectively.
The coupling constants in these folded systems are not
independent as shown in (\ref{foldrel1}).

%%%%%%%%%%%%%%%%%%%%%%%%%%%%
%  2.2.1                   %
%%%%%%%%%%%%%%%%%%%%%%%%%%%%
\subsubsection{Ruijsenaars-Calogero Systems}

The first example to be  discussed is the `discrete' analogue of the
Calogero systems \cite{Cal}, to be called the Ruijsenaars-Calogero systems,
which were introduced by van Diejen for the classical root systems only
\cite{vanDiejen1,vanDiejen2}.
The original Calogero systems \cite{Cal} have the rational
($1/$(distance)${}^2$) potential plus the harmonic  confining potential,
having two coupling constants $g$ and $\omega$ for the systems based on
the simply-laced root systems, $A$ and $D$, and three couplings $\omega$
and $g_L$ for the long roots and $g_S$ for the short roots in
the $B$ and $C$ root systems.

Two varieties (deformation) of `discrete' Calogero systems are known.
The first has two (three for the non-simply-laced root systems) coupling
constants $g$ ($g_L$ and $g_S$) and $a$ which corresponds to $\omega$ in
the Calogero systems.
The second has three (four for the non-simply-laced root systems)
coupling constants $g$ ($g_L$ and $g_S$) and $a,\,b$ both of which
correspond to $\omega$.
The integrability (classical and quantum) of these systems was discussed
by van Diejen in some detail \cite{vanDiejen1,vanDiejen2}.
The former can be considered as a limiting case ($b\to\infty$) of the
latter.

%%%%%%%%%%%%%%%%%%%%%%%%%%%%
%  2.2.1.1                 %
%%%%%%%%%%%%%%%%%%%%%%%%%%%%
\paragraph{Linear Confining Potential Case}~

\noindent
The dynamical system is defined by giving  the explicit forms of the
elementary potential functions $v$ and $w$.
For the simply-laced root systems $A$ and $D$ they are:
\begin{eqnarray}
  \hspace*{-3.7cm}
  A,\ D\ :\ \ v(x)=1-i\,\frac{\,g\,}{x}\,,\quad
  w(x)=1+i\,\frac{\,x\,}{a}\,,
  \label{adminC}
\end{eqnarray}
in which $a$ and $g$ are real positive coupling constants.
For the non-simply-laced root systems $B$, $C$ and $\widetilde{BC}$,
we have:
\begin{eqnarray}
  B&:& v(x)=1-i\,\frac{g_L}{x}\,,\quad
  w(x)=\Bigl(1+i\,\frac{\,x\,}{a}\Bigr)\Bigl(1-i\,\frac{g_S}{2x}\Bigr)^{\!\!2},
  \label{bminC}\\
  C&:& v(x)=1-i\,\frac{g_S}{x}\,,\quad
  w(x)=\Bigl(1+i\,\frac{\,x\,}{a}\Bigr)\Bigl(1-i\,\frac{g_L}{x}\Bigr)\,,
  \label{cminC}\\
  \widetilde{BC}&:& v(x)=1-i\,\frac{g_0}{x}\,,\quad
  w(x)=\Bigl(1+i\,\frac{\,x\,}{a}\Bigr)\Bigl(1-i\,\frac{g_1}{x}\Bigr)
  \Bigl(1-i\,\frac{g_2}{x}\Bigr)\,,
  \label{bctminC}
\end{eqnarray}
in which $a,g_L,g_S,g_0,g_1,g_2$ are {\em
independent\/} real positive coupling constants.
Normalisation of the coupling constants is chosen such
that they reduce to those of the Calogero models in the
small coupling limits discussed below. The $D$ model is
obtained from the $B$ model by $g_L=g$ and $g_S=0$.
The $B$ and $C$ models are special cases of the
$\widetilde{BC}$ model. In contrast to the Calogero case,
those based on the $B$ and $C$ systems are different. 
The difference of the length of the roots is immaterial, 
since it can be absorbed by the coupling constants normalisation. 
It is rather elementary to derive the forms of the Hamiltonians of 
the $B$ and $C$ systems from those of the $D$ and $A$ systems by 
{\em folding\/} (see (\ref{DtoB}) and (\ref{AtoC})).
In all these cases the `potential' $P$ (\ref{potP}) grows linearly in
$|q|$ as $|q|\to\infty$.
Except for the $\widetilde{BC}$ case, there are simple identities :
\(
  \sum_{j}\left(V_j(q)+V_j^*(q)\right)=\mbox{const}.
\)

In the limit of small coupling constants,
namely, by recovering the deformation parameter $\beta'$,
\begin{equation}
  (p_j,\frac{1}{a},g,g_L,g_S,g_0,g_1,g_2)\rightarrow
  \beta'\,(p_j,\frac{1}{a},g,g_L,g_S,g_0,g_1,g_2),
  \label{smallcoupling}
\end{equation}
and taking $\beta'\rightarrow 0$ limit,
the Hamiltonian (\ref{Ham}) tends to that of the
corresponding classical Calogero system (\ref{HCS}) with
$\omega=\frac{1}{a}$ ($\widetilde{BC}_r$ tends to $C_r$ with
$g_S=g_0$, $g_L=g_1+g_2$)
\begin{equation}
  \frac{1}{\beta^{\prime\,2}}H(p,q)\to H_{\rm Calogero}(p,q)+\mbox{const}.
  \label{callim}
\end{equation}

It is interesting to note that the equations determining the equilibrium
(\ref{equeq}), in general, can be cast in a form which looks similar to
the {\em Bethe ansatz\/} equation. For the elementary potential
(\ref{adminC})--(\ref{bctminC}), the equilibrium positions $\{\bar{q}_j\}$
are determined by:
\begin{eqnarray}
  A_{r-1}&:&\prod_{k=1\atop k\neq j}^r
  {\bar{q}_j-\bar{q}_k-ig\over{\bar{q}_j-\bar{q}_k+ig}}=
  {a-i\bar{q}_j\over{a+i\bar{q}_j}}\,,
  \label{rclAeq}\\
  B_r&:&\prod_{k=1\atop k\neq j}^r
  {\bar{q}_j-\bar{q}_k-ig_L\over{\bar{q}_j-\bar{q}_k+ig_L}}\,
  {\bar{q}_j+\bar{q}_k-ig_L\over{\bar{q}_j+\bar{q}_k+ig_L}}=
  {a-i\bar{q}_j\over{a+i\bar{q}_j}}
  \left({2\bar{q}_j+ig_S\over{2\bar{q}_j-ig_S}}\right)^{\!\!2}\!,
  \label{rclBeq}\\
  C_r&:&\prod_{k=1\atop k\neq j}^r
  {\bar{q}_j-\bar{q}_k-ig_S\over{\bar{q}_j-\bar{q}_k+ig_S}}\,
  {\bar{q}_j+\bar{q}_k-ig_S\over{\bar{q}_j+\bar{q}_k+ig_S}}=
  {a-i\bar{q}_j\over{a+i\bar{q}_j}}\,
  {\bar{q}_j+ig_L\over{\bar{q}_j-ig_L}}\,,
  \label{rclCeq}\\
  \widetilde{BC}_r&:&\prod_{k=1\atop k\neq j}^r
  {\bar{q}_j-\bar{q}_k-ig_0\over{\bar{q}_j-\bar{q}_k+ig_0}}\,
  {\bar{q}_j+\bar{q}_k-ig_0\over{\bar{q}_j+\bar{q}_k+ig_0}}=
  {a-i\bar{q}_j\over{a+i\bar{q}_j}}\,
  {\bar{q}_j+ig_1\over{\bar{q}_j-ig_1}}\,
  {\bar{q}_j+ig_2\over{\bar{q}_j-ig_2}}\,,
  \label{rclBCteq}\\
  D_r&:&\prod_{k=1\atop k\neq j}^r
  {\bar{q}_j-\bar{q}_k-ig\over{\bar{q}_j-\bar{q}_k+ig}}\,
  {\bar{q}_j+\bar{q}_k-ig\over{\bar{q}_j+\bar{q}_k+ig}}=
  {a-i\bar{q}_j\over{a+i\bar{q}_j}}\,.
  \label{rclDeq}
\end{eqnarray}
In the above small coupling limits, these equations reduce to
(\ref{herdet})--(\ref{lagdetD}).
Thus the {\em Bethe ansatz\/}-like equations (\ref{rclAeq})--(\ref{rclDeq})
would give deformation of the Hermite and Laguerre polynomials,
as we will see in section \ref{sec:H.1} and \ref{sec:L.1}.
They are not the so-called $q$-{\em deformed\/} Hermite or Laguerre
polynomials \cite{And-Ask-Roy}.

%%%%%%%%%%%%%%%%%%%%%%%%%%%%
%  2.2.1.2                 %
%%%%%%%%%%%%%%%%%%%%%%%%%%%%
\paragraph{Quadratic Confining Potential Case}~

\noindent
In this case  the elementary potential function $v$ is the same as
before, but $w$ is different.
For the simply-laced root systems $A$ and $D$, the elementary potential
functions are:
\begin{eqnarray}
  A,\ D\ :\ \ v(x)=1-i\,\frac{\,g\,}{x},\quad
  w(x)=\Bigl(1+i\,\frac{\,x\,}{a}\Bigr)\Bigl(1+i\,\frac{\,x\,}{b}\Bigr),
  \quad (a,\, b, \, g>0).
  \label{ADVform}
\end{eqnarray}
For the non-simply-laced root systems $B$, $C$ and $\widetilde{BC}$, we
have  ($g_L$, $g_S$, $g_0$, $g_1$, $g_2>0$):
\begin{eqnarray}
  B&:& v(x)=1-i\,\frac{g_L}{x},\quad
  w(x)=\Bigl(1+i\,\frac{\,x\,}{a}\Bigr)\Bigl(1+i\,\frac{\,x\,}{b}\Bigr)
  \Bigl(1-i\,\frac{g_S}{2x}\Bigr)^{\!\!2},
  \label{Bpot}\\
  C&:& v(x)=1-i\,\frac{g_S}{x},\quad
  w(x)=\Bigl(1+i\,\frac{\,x\,}{a}\Bigr)\Bigl(1+i\,\frac{\,x\,}{b}\Bigr)
  \Bigl(1-i\,\frac{g_L}{x}\Bigr),
  \label{Cpot}\\
  \widetilde{BC}&:&v(x)=1-i\,\frac{g_0}{x},\quad
  w(x)=\Bigl(1+i\,\frac{\,x\,}{a}\Bigr)\Bigl(1+i\,\frac{\,x\,}{b}\Bigr)
  \Bigl(1-i\,\frac{g_1}{x}\Bigr)\Bigl(1-i\,\frac{g_2}{x}\Bigr).
  \label{BCpot}
\end{eqnarray}
The $D$ model can be considered as a special case of the
$B$ model by $g_L=g$ and $g_S=0$.
The $B$ and $C$ models are special cases of
$\widetilde{BC}$ model. As in the previous case, the forms
of the elementary potential function
$w$ for $B$ and $C$ systems are determined from those of
the $D$ and $A$ systems by folding (\ref{DtoB}),
(\ref{AtoC}). In all these cases the `potential' $P$
(\ref{potP}) grows quadratically in $|q|$ as
$|q|\to\infty$. The small coupling limit (\ref{smallcoupling}) 
(and $\frac{1}{b}\rightarrow\frac{\beta'}{b}$) gives the same
classical Calogero systems as before (\ref{callim}),
(\ref{HCS}) with $\omega=\frac{1}{a}+\frac{1}{b}$.

The equations (\ref{equeq}) determining the equilibrium positions
$\{\bar{q}_j\}$ for the elementary potential
(\ref{ADVform})--(\ref{BCpot}) are expressed in a form similar to the
{\em Bethe ansatz\/} equation:
\begin{eqnarray}
  A_{r-1}&:&\prod_{k=1\atop k\neq j}^r
  {\bar{q}_j-\bar{q}_k-ig\over{\bar{q}_j-\bar{q}_k+ig}}=
  {a-i\bar{q}_j\over{a+i\bar{q}_j}}{b-i\bar{q}_j\over{b+i\bar{q}_j}}\,,
  \label{rcqAeq}\\
  B_r&:&\prod_{k=1\atop k\neq j}^r
  {\bar{q}_j-\bar{q}_k-ig_L\over{\bar{q}_j-\bar{q}_k+ig_L}}\,
  {\bar{q}_j+\bar{q}_k-ig_L\over{\bar{q}_j+\bar{q}_k+ig_L}}=
  {a-i\bar{q}_j\over{a+i\bar{q}_j}}\,
  {b-i\bar{q}_j\over{b+i\bar{q}_j}}
  \left({2\bar{q}_j+ig_S\over{2\bar{q}_j-ig_S}}\right)^{\!\!2}\!,
  \label{rcqBeq}\\
  C_r&:&\prod_{k=1\atop k\neq j}^r
  {\bar{q}_j-\bar{q}_k-ig_S\over{\bar{q}_j-\bar{q}_k+ig_S}}\,
  {\bar{q}_j+\bar{q}_k-ig_S\over{\bar{q}_j+\bar{q}_k+ig_S}}=
  {a-i\bar{q}_j\over{a+i\bar{q}_j}}\,
  {b-i\bar{q}_j\over{b+i\bar{q}_j}}\,
  {\bar{q}_j+ig_L\over{\bar{q}_j-ig_L}}\,,
  \label{rcqCeq}\\
  \widetilde{BC}_r&:&\prod_{k=1\atop k\neq j}^r
  {\bar{q}_j-\bar{q}_k-ig_0\over{\bar{q}_j-\bar{q}_k+ig_0}}\,
  {\bar{q}_j+\bar{q}_k-ig_0\over{\bar{q}_j+\bar{q}_k+ig_0}}=
  {a-i\bar{q}_j\over{a+i\bar{q}_j}}\,
  {b-i\bar{q}_j\over{b+i\bar{q}_j}}\,
  {\bar{q}_j+ig_1\over{\bar{q}_j-ig_1}}\,
  {\bar{q}_j+ig_2\over{\bar{q}_j-ig_2}}\,,\ \
  \label{rcqBCteq}\\
  D_r&:&\prod_{k=1\atop k\neq j}^r
  {\bar{q}_j-\bar{q}_k-ig\over{\bar{q}_j-\bar{q}_k+ig}}\,
  {\bar{q}_j+\bar{q}_k-ig\over{\bar{q}_j+\bar{q}_k+ig}}=
  {a-i\bar{q}_j\over{a+i\bar{q}_j}}\,
  {b-i\bar{q}_j\over{b+i\bar{q}_j}}\,.
  \label{rcqDeq}
\end{eqnarray}
They define another type of deformation of the Hermite and Laguerre
polynomials, since the small coupling limit of the above Bethe
ansatz-like equations gives the same equations as before
(\ref{herdet})--(\ref{lagdetD}), determining the zeros of the Hermite
and Laguerre polynomials, with $\omega=\frac{1}{a}+\frac{1}{b}$.
These will be discussed in sections \ref{sec:H.2} and \ref{sec:L.2}.

%%%%%%%%%%%%%%%%%%%%%%%%%%%%
%  2.2.2                   %
%%%%%%%%%%%%%%%%%%%%%%%%%%%%
\subsubsection{Ruijsenaars-Sutherland Systems}

The discrete analogue of the Sutherland systems \cite{Sut}, to be called
the Ruijsenaars-Sutherland systems, was introduced originally by
Ruijsenaars and Schneider \cite{Ruij-Sch} for the $A$ type root system.
The quantum eigenfunctions of the $A$ type Ruijsenaars-Sutherland
systems are called Macdonald polynomials \cite{Macdonald}, which are a
one-parameter deformation ($q$-deformation) of the Jack polynomials
\cite{Jack}.
Here we will discuss the Ruijsenaars-Sutherland systems for all the
classical root systems, $A$, $B$, $C$, $D$ and $BC$ \cite{vanDiejen2}.
The structure of the functions $\{V_j(q)\}$, (\ref{AVform}) and
(\ref{otherVform}) are the same as in  the Ruijsenaars-Calogero systems,
but the elementary potential functions $v$ and $w$ are trigonometric
instead of rational. Because of the identity
$\sum_{j=1}^r\left\{V_j(q)+V_j^*(q)\right\}=\mbox{const}.$,
the Hamiltonian (\ref{Ham}) could be replaced by a simpler one
(\ref{Ham2}).

The elementary potential functions $v$ and $w$ are:
\begin{eqnarray}
  A,\,D&:& v(x)=1-i\tanh g\cot x, \quad w(x)=1,
  \label{sutADdef}\\
  B&:& v(x)=1-i\tanh g_L\cot x,
  \quad w(x)=(1-i\tanh\frac{g_S}{2}\cot x)^2,
  \label{sutBdef}\\
  B'&:& v(x)=1-i\tanh g_L\cot x,
  \quad w(x)=1-i\tanh g_S\cot x,
  \label{sutB'def}\\
  C&:& v(x)=1-i\tanh g_S\cot x,
  \quad w(x)=1-i\tanh 2g_L\cot 2x,
  \label{sutCdef}\\
  C'&:& v(x)=1-i\tanh g_S\cot x,
  \quad w(x)=(1-i\tanh g_L\cot 2x)^2,
  \label{sutC'def}\\
  B'C&:& v(x)=1-i\tanh g_M\cot x,\nonumber\\
  &&\qquad\qquad w(x)=(1-i\tanh g_S\cot x)(1-i\tanh 2g_L\cot 2x),
  \label{sutBCdef}
\end{eqnarray}
with similar coupling constant notation as in the rational cases.
Normalisation of the coupling constants is chosen such that they 
reduce to those of the Sutherland models in the small coupling limits 
discussed below. The $D$ model is the special case of the $B$ 
model by $g_L=g$ and $g_S=0$.
The $B'$ and $C$ models are special cases of the $B'C$
model. As in the rational cases, the forms of the elementary potential 
function $w$ for the $B$ and $C$ systems are determined from those of 
the $D$ and $A$ systems by folding (\ref{DtoB}), (\ref{AtoC}).

The original Sutherland models are obtained in the limit in
which all the coupling constant(s) become infinitesimally small.
By recovering the deformation parameter $\beta'$,
\begin{equation}
  (p_j,g,g_L,g_M,g_S)\rightarrow
  \beta'\,(p_j,g,g_L,g_M,g_S),
  \label{smallcoupling2}
\end{equation}
and taking $\beta'\rightarrow 0$ limit,
the Hamiltonian (\ref{Ham}) tends to that of the
corresponding classical Sutherland system (\ref{HCS})
\begin{equation}
  \frac{1}{\beta^{\prime\,2}}H(p,q)\to H_{\rm Sutherland}(p,q)+\mbox{const}.
  \label{sutlim}
\end{equation}

In `strong' coupling limits, $g$, $g_L$, $g_M$, $g_S$ $\to+\infty$,
the elementary potential functions $v$ and $w$ take simple forms:
\begin{equation}
  v(x)\to 1-i\cot x,\qquad
  w(x)\to1, \quad 1-i\cot x,\quad  (1-i\cot x)^2,\quad\mbox{etc}.
  \label{stronglim}
\end{equation}
The deformed polynomials take simple forms in this limit as we will see in
section \ref{sec:J}.

The equations (\ref{equeq}) determining the equilibrium positions
$\{\bar{q}_j\}$ for the elementary potential
(\ref{sutADdef})--(\ref{sutBCdef}) are expressed in a form similar to the
{\em Bethe ansatz\/} equation:
\begin{eqnarray}
  A_{r-1}&\!\!:\!\!&\prod_{k=1\atop k\neq j}^r
  {\tan(\bar{q}_j-\bar{q}_k)-i\tanh g\over
  {\tan(\bar{q}_j-\bar{q}_k)+i\tanh g}}=
  1,
  \label{rsAeq}\\
  B_r&\!\!:\!\!&\prod_{k=1\atop k\neq j}^r
  {\tan(\bar{q}_j-\bar{q}_k)-i\tanh g_L\over
  {\tan(\bar{q}_j-\bar{q}_k)+i\tanh g_L}}\,
  {\tan(\bar{q}_j+\bar{q}_k)-i\tanh g_L\over
  {\tan(\bar{q}_j+\bar{q}_k)+i\tanh g_L}}=
  \left({\tan\bar{q}_j+i\tanh\frac{g_S}{2}\over
  {\tan\bar{q}_j-i\tanh\frac{g_S}{2}}}\right)^{\!\!2}\!\!,
  \label{rsBeq}\\
  B'_r&\!\!:\!\!&\prod_{k=1\atop k\neq j}^r
  {\tan(\bar{q}_j-\bar{q}_k)-i\tanh g_L\over
  {\tan(\bar{q}_j-\bar{q}_k)+i\tanh g_L}}\,
  {\tan(\bar{q}_j+\bar{q}_k)-i\tanh g_L\over
  {\tan(\bar{q}_j+\bar{q}_k)+i\tanh g_L}}=
  {\tan\bar{q}_j+i\tanh g_S\over
  {\tan\bar{q}_j-i\tanh g_S}}\,,
  \label{rsBpeq}\\
  C_r&\!\!:\!\!&\prod_{k=1\atop k\neq j}^r
  {\tan(\bar{q}_j-\bar{q}_k)-i\tanh g_S\over
  {\tan(\bar{q}_j-\bar{q}_k)+i\tanh g_S}}\,
  {\tan(\bar{q}_j+\bar{q}_k)-i\tanh g_S\over
  {\tan(\bar{q}_j+\bar{q}_k)+i\tanh g_S}}=
  {\tan 2\bar{q}_j+i\tanh 2g_L\over
  {\tan 2\bar{q}_j-i\tanh 2g_L}}\,,
  \label{rsCeq}\\
  C'_r&\!\!:\!\!&\prod_{k=1\atop k\neq j}^r
  {\tan(\bar{q}_j-\bar{q}_k)-i\tanh g_S\over
  {\tan(\bar{q}_j-\bar{q}_k)+i\tanh g_S}}\,
  {\tan(\bar{q}_j+\bar{q}_k)-i\tanh g_S\over
  {\tan(\bar{q}_j+\bar{q}_k)+i\tanh g_S}}=
  \left({\tan 2\bar{q}_j+i\tanh g_L\over
  {\tan 2\bar{q}_j-i\tanh g_L}}\right)^{\!\!2}\!\!,\qquad
  \label{rsCpeq}\\
  B'C_r&\!\!:\!\!&\prod_{k=1\atop k\neq j}^r
  {\tan(\bar{q}_j-\bar{q}_k)-i\tanh g_M\over
  {\tan(\bar{q}_j-\bar{q}_k)+i\tanh g_M}}\,
  {\tan(\bar{q}_j+\bar{q}_k)-i\tanh g_M\over
  {\tan(\bar{q}_j+\bar{q}_k)+i\tanh g_M}}\nonumber\\
  &&\qquad\qquad\qquad\qquad\qquad\qquad\qquad=
  {\tan\bar{q}_j+i\tanh g_S\over
  {\tan\bar{q}_j-i\tanh g_S}}\,
  {\tan2\bar{q}_j+i\tanh 2g_L\over
  {\tan2\bar{q}_j-i\tanh 2g_L}}\,,
  \label{rsBCeq}\\
  D_r&\!\!:\!\!&\prod_{k=1\atop k\neq j}^r
  {\tan(\bar{q}_j-\bar{q}_k)-i\tanh g\over
  {\tan(\bar{q}_j-\bar{q}_k)+i\tanh g}}\,
  {\tan(\bar{q}_j+\bar{q}_k)-i\tanh g\over
  {\tan(\bar{q}_j+\bar{q}_k)+i\tanh g}}=
  1.
  \label{rsDeq}
\end{eqnarray}
{}From the property mentioned after (\ref{equeq}) and the
fact that (\ref{rsAeq})--(\ref{rsDeq}) are the equations of
$\{\tan\bar{q}_j\}$, we can restrict
$\bar{q}_j$ to $0\leq\bar{q}_j\leq {\pi}/{2}$ (except
for the $A_{r-1}$ case). In the small coupling limit, these
equations  (\ref{rsAeq})--(\ref{rsDeq}) tend to
(\ref{chedet})--(\ref{jacobidefD}). Thus the {\em Bethe
ansatz\/}-like equations (\ref{rsAeq})--(\ref{rsDeq})
would give deformation of the Chebyshev and Jacobi
polynomials as we will see in section \ref{sec:J}.

%%%%%%%%%%%%%%%%%%%%%%%%%%%%
%  2.3                     %
%%%%%%%%%%%%%%%%%%%%%%%%%%%%
\subsection{Orthogonal Polynomials and Three Term Recurrences}

It is well known that orthogonal polynomials satisfy
three-term recurrence \cite{szego,Chihara} and conversely
a sequence of polynomials satisfying three-term recurrence
are orthogonal with respect to certain inner product
with some weight function.
Here we will introduce appropriate notation by taking the
classical orthogonal polynomials as examples.

Let $\{f_n(x)\}_{n=0}^{\infty}$ be a sequence of orthogonal polynomials
with real coefficients. That is $f_n(x)$ is a degree $n$
polynomial in $x$ and  they are mutually orthogonal
$(f_n,f_m)=h_n\delta_{n,m}$ ($h_n>0$) with  respect to an
(positive definite) inner product
$(f,g)=\int f(x)g(x){\mathbf w}(x)dx$
(${\mathbf w}(x)$ is a weight function).
Let $f_n^{\rm monic}(x)$ be a monic one,
$f_n(x)=c_nf_n^{\rm monic}(x)=c_n(x^n+\cdots)$.
Then $f_n^{\rm monic}(x)$ satisfies three-term recurrence :
\begin{equation}
  f_{n+1}^{\rm monic}(x)-(x-a_n)f_n^{\rm monic}(x)
  +b_nf_{n-1}^{\rm monic}(x)=0\,,\quad(n\geq 0)\,,
  \label{3termrec}
\end{equation}
where we have set $f^{\rm monic}_{-1}(x)=0$\,%
\footnote{
Hereafter we adopt the convention $f_{-1}(x)=0$ and
$f_0(x)=1$ for all the polynomials in this paper.}, and
$a_n$ ($n\geq 0$) and $b_n$ ($n\geq 1$, $b_0$ is unnecessary, $b_n>0$)
are real numbers.
The constants $a_n$, $b_n$ and $h_n$ are given by
\begin{equation}
  a_n=\frac{(xf_n(x),f_n(x))}{(f_n(x),f_n(x))}\,,\quad
  b_n=\frac{c_{n-1}^2}{c_n^2}
  \frac{(f_n(x),f_n(x))}{(f_{n-1}(x),f_{n-1}(x))}\,,\quad
  h_n=(1,1)c_n^2\prod_{j=1}^nb_j\,.
  \label{abh}
\end{equation}
Namely $f_n(x)$ satisfies the three-term recurrence
\begin{equation}
  \frac{c_n}{c_{n+1}}f_{n+1}(x)-(x-a_n)f_n(x)
  +b_n\frac{c_n}{c_{n-1}}f_{n-1}(x)=0\,,\quad (n\geq 0).
\end{equation}
For $a_n=0$ ($n\geq 0$) case, $f_n(x)$ has a definite parity,
$f_n(-x)=(-1)^nf_n(x)$, and the constant term of the even polynomial is
\begin{eqnarray}
  f_{2n}(0)=(-1)^nc_{2n}\prod_{j=1}^nb_{2j-1}.
  \label{originval}
\end{eqnarray}

Conversely, if $\{f_n(x)\}$ is defined by the three-term recurrence
(\ref{3termrec}), namely, real numbers $a_n$ ($n\geq 0$),
$b_n$ ($n\geq 1$, $b_n>0$) and $c_n$ ($n\geq 0$, $c_n\neq 0$) are
given, then $\{f_n(x)\}$ is a sequence of orthogonal polynomials with
respect to some (positive definite) inner product $(\cdot\,,\cdot\,)$.

\bigskip
In the rest of this subsection we summarise the three-term recurrence,
generating functions, differential equations, {\em etc\/} for the
Hermite, Laguerre and Jacobi polynomials for later comparison with the
corresponding quantities of the deformed polynomials.

The Hermite polynomials $H_n(x)$ (\ref{hermdef}) are orthogonal
with respect to the inner product 
$(f,g)$ $=\int_{-\infty}^{\infty}f(x)g(x)e^{-x^2}dx$,
($h_n=2^nn!\sqrt{\pi}$)
and satisfy the three-term recurrence (\ref{3termrec}) with
\begin{eqnarray}
  && a_n=0,\qquad b_n=\frac{n}{2},\qquad c_n=2^n,
  \label{herrec}\\ 
  &&H_{n+1}(x)-2xH_n(x)+2nH_{n-1}(x)=0.
\end{eqnarray}
The generating function and orthogonality are
\begin{equation}
  G(t,x)\eqdef\sum_{n=0}^{\infty}\frac{t^n}{n!}H_n(x)=e^{-t^2+2xt},\quad
  (G(t,x),G(s,x))=\sqrt{\pi}e^{2ts}.
  \label{gf_H}
\end{equation}

The Laguerre polynomials $L^{(\alpha)}_n(x)$ (\ref{lagdef}) are
orthogonal  with respect to the inner product
$(f,g)=\int_0^{\infty}f(x)g(x)x^{\alpha}e^{-x}dx$,
($h_n(\alpha)=\Gamma(\alpha+n+1)/n!$ , ${\rm Re}\,\alpha>-1$)
and satisfy the three-term recurrence (\ref{3termrec}) with
\begin{eqnarray}
  && a_n=2n+1+\alpha,\qquad b_n=n(n+\alpha),\qquad
  c_n={(-1)^n}/{n!},
  \label{Lagrec}\\
  && (n+1)L_{n+1}^{(\alpha)}(x)+(x-(2n+\alpha+1))L_n^{(\alpha)}(x)
  +(n+\alpha)L_{n-1}^{(\alpha)}(x)=0.
\end{eqnarray}

The Jacobi polynomials $P_n^{(\alpha,\beta)}(x)$ (\ref{Pdef}) are
orthogonal with respect to
$(f,g)=\int_{-1}^{1}f(x)$ $g(x)(1-x)^{\alpha}(1+x)^{\beta}dx$,
($h_n(\alpha,\beta)=\frac{2^{\alpha+\beta+1}}{\alpha+\beta+2n+1}
\frac{\Gamma(\alpha+n+1)\Gamma(\beta+n+1)}{n!\Gamma(\alpha+\beta+n+1)}$)
and satisfy the three-term recurrence (\ref{3termrec}) with
\begin{eqnarray}
  &&a_n=\frac{\beta^2-\alpha^2}{d_{2n}\,d_{2n+2}}\,,\quad
  b_n=4n(n+\alpha)(n+\beta)\times
  \frac{d_n}{d_{2n-1}\,d_{2n}^{\ 2}\,d_{2n+1}}\,,
  \label{Jacrec}\\
  &&c_n=2^{-n}\schoose{\alpha+\beta+2n}{n},\quad
  d_m=\alpha+\beta+m,\qquad\qquad\qquad\quad \\
  &&2(n+1)(\alpha+\beta+n+1)(\alpha+\beta+2n)P_{n+1}^{(\alpha,\beta)}(x)
  \n
  &&\quad
  -(\alpha+\beta+2n+1)\Bigl((\alpha+\beta+2n)(\alpha+\beta+2n+2)x
  +\alpha^2-\beta^2\Bigr)P_n^{(\alpha,\beta)}(x)\n
  &&\qquad
  +2(\alpha+n)(\beta+n)(\alpha+\beta+2n+2)P_{n-1}^{(\alpha,\beta)}(x)=0\,.
\end{eqnarray}

The Gegenbauer polynomial $C^{(\alpha+\frac12)}_n(x)$ (\ref{Gegendef})
is a special case of the Jacobi polynomial
$C^{(\alpha+\frac12)}_n(x)=\schoose{2\alpha+n}{n}\schoose{\alpha+n}{n}^{-1}
P^{(\alpha,\alpha)}_n(x)$.
The Chebyshev polynomial of the first kind $T_n(x)$ is also a special case of
the Jacobi polynomial
$T_n(x)=2^{-1}\schoose{2n-1}{n}^{-1}P^{(-\frac12,-\frac12)}_n(x)$.
The Legendre polynomial is $P_n(x)=P_n^{(0,0)}(x)$

The differential equations of the Hermite, Laguerre and Jacobi
polynomials are
\begin{eqnarray}
  &&\sfrac{d^2}{dx^2}H_n(x)-2x\sfrac{d}{dx}H_n(x)+2nH_n(x)=0\,,
  \label{diffeqH}\\
  &&x\sfrac{d^2}{dx^2}L_n^{(\alpha)}(x)
  +(\alpha+1-x)\sfrac{d}{dx}L_n^{(\alpha)}(x)+nL_n^{(\alpha)}(x)=0\,,
  \label{diffeqL}\\
  &&(1-x^2)\sfrac{d^2}{dx^2}P_n^{(\alpha,\beta)}(x)
  +\Bigl(\beta-\alpha-(\alpha+\beta+2)x\Bigr)
  \sfrac{d}{dx}P_n^{(\alpha,\beta)}(x)\n
  &&\qquad\qquad\qquad\qquad\qquad\qquad
  +n(n+\alpha+\beta+1)P_n^{(\alpha,\beta)}(x)=0\,.
  \label{diffeqJ}
\end{eqnarray}

%%%%%%%%%%%%%%%%%%%%%%%%%%%%%%%%%%%%%%%%%%%%%%%%%%%%%%%%%%%%%%%
%                                                             %
%  3. Deformations of the Hermite Polynomial                  %
%                                                             %
%%%%%%%%%%%%%%%%%%%%%%%%%%%%%%%%%%%%%%%%%%%%%%%%%%%%%%%%%%%%%%%
\section{Deformation of the Hermite Polynomial}
\label{sec:H}
\setcounter{equation}{0}

%%%%%%%%%%%%%%%%%%%%%%%%%%%%
%  3.1                     %
%%%%%%%%%%%%%%%%%%%%%%%%%%%%
\subsection{Linear Confining Potential Case (1 Parameter Deformation)}
\label{sec:H.1}

For the solution $\{\bar{q}_j\}$ of the $A_{r-1}$ equation (\ref{rclAeq}),
let us define
\begin{equation}
  \bar{q}_j=\sqrt{ag}\,y_j\,,\quad \delta=\frac{g}{a}\,,
  \label{para_rclA}
\end{equation}
and introduce a degree $r$ polynomial in $x$ having zeros at $\{y_j\}$:
\begin{equation}
  H_r(x,\delta)\eqdef 2^r\prod_{j=1}^r(x-y_j)\,.
\end{equation}
It is a deformation of the Hermite polynomial (\ref{hermdef}) such that
\begin{equation}
  \lim_{\delta\to 0}H_r(x,\delta)=H_r(x).
\end{equation}
If $\{\bar{q}_j\}$ is a solution of (\ref{rclAeq}), so is
$\{-\bar{q}_j\}$, which would imply that the deformed polynomial
$H_r(x,\delta)$ has a definite parity
\begin{equation}
  H_r(-x,\delta)=(-1)^rH_r(x,\delta),
\end{equation}
as with the original Hermite polynomial $H_r(-x)=(-1)^rH_r(x)$.

The equation for the equilibrium (\ref{rclAeq}) can be written as (we
replace $r$ by $n$)
\begin{equation}
  \prod_{k=1}^n
  {y_j-y_k-i\dr\over{y_j-y_k+i\dr}}=
  {y_j+i\frac{1}{\dr}\over{y_j-i\frac{1}{\dr}}}\,.
  \label{rclAeq2}
\end{equation}
{}From this equation, we obtain the following functional equation
for $H_n(x,\delta)$ ($n\geq 1$),
\begin{equation}
  \Bigl(x+i\frac{1}{\dr}\Bigr)H_n(x+i\dr,\delta)
  -\Bigl(x-i\frac{1}{\dr}\Bigr)H_n(x-i\dr,\delta)
  =2iA_nH_n(x,\delta),
  \label{feq_rclA}
\end{equation}
because the LHS is $i$ times a degree $n$ polynomial in $x$
with real coefficients which vanishes at $x=y_j$.
Here $A_n=A_n(\delta)$ is a real constant.
This functional equation contains all the information of the equilibrium.
The number of unknown coefficients (coefficient of $x^k$ term of
$H_n(x,\delta)$ ($k=0,1,\ldots,n-1$) and $A_n$) and the number of
equations (coefficient of $x^k$ term of (\ref{feq_rclA})
($k=0,1,\ldots,n$)) are both $n$.
The constant $A_n$ is determined by the coefficient of $x^n$ term of this
equation,
\begin{equation}
  A_n=\frac{1}{\dr}(1+n\delta\,).
  \label{feq_rclA_coef}
\end{equation}
The functional equation (\ref{feq_rclA}) can be written as a difference
equation,
\begin{equation}
  D_{x,\frac12i\dr}^2\,H_n(x,\delta)
  -2xD_{x,i\dr}\,H_n(x,\delta) +2nH_n(x,\delta)=0,
\end{equation}
where $D_{x,h}$ is a (central) difference operator,
\begin{equation}
  D_{x,h}f(x)=\frac{f(x+h)-f(x-h)}{2h}\,.
\end{equation}
In the $\delta\to 0$ limit, (\ref{feq_rclA}) reduces
to the differential equation of the Hermite polynomial (\ref{diffeqH}).

The uniqueness (up to normalisation) of the solution of the functional
equation (\ref{feq_rclA}) is easily shown (Proposition \ref{prop:1}).
Therefore it is sufficient to construct one solution of
(\ref{feq_rclA}) explicitly.
This is done by using the three-term recurrence (see Appendix).
The result is as follows;
The functional equation (\ref{feq_rclA}) implies
that the deformed Hermite polynomial $H_n(x,\delta)$ satisfies the
three-term recurrence (\ref{3termrec}) with
\begin{eqnarray}
  &&a_n=0,\qquad
  b_n=\frac{n}{2}\Bigl(1+\frac{n-1}{2}\delta\Bigr),\qquad c_n=2^n,
  \label{rec_rclA}\\
  &&H_{n+1}(x,\delta)-2xH_n(x,\delta)
  +\Bigl(2n+n(n-1)\delta\Bigr)H_{n-1}(x,\delta)=0.
  \label{rec_rclA2}
\end{eqnarray}
Since $\delta$ is positive in this case (\ref{para_rclA}), $b_n$ is
also positive. Therefore $H_n(x,\delta)$ is a set of orthogonal
polynomials with respect to some positive definite inner product.

Here we present another derivation of this three-term recurrence.
Let us consider the generating function
\begin{equation}
  G(t,x,\delta)=\sum_{n=0}^{\infty}\frac{t^n}{n!}H_n(x,\delta),
\end{equation}
which satisfies
\begin{equation}
  \Bigl((1+\delta\,t^2)\frac{\partial}{\partial t}+2(t-x)\Bigr)
  G(t,x,\delta)=0,
\end{equation}
as a consequence of the three-term recursion (\ref{rec_rclA2}).
This linear differential equation with the initial condition
$G(0,x,\delta)=1$ can be easily solved and we obtain
\begin{equation}
  G(t,x,\delta)=\frac{\exp(2x\frac{\arctan\dr\,t}{\dr})}
  {(1+\delta\,t^2)^{\frac{1}{\delta}}}.
  \label{G_ans}
\end{equation}
In the $\delta\to 0$ limit, this generating function tends to that
of the Hermite polynomial (\ref{gf_H}),
\begin{equation}
  \lim_{\delta\to 0}G(t,x,\delta)=e^{-t^2}e^{2xt}=e^{-t^2+2xt}=G(t,x).
\end{equation}
The functional equation of $G(t,x,\delta)$ is obtained from
(\ref{feq_rclA}):
\begin{equation}
  \Bigl(x+i\frac{1}{\dr}\Bigr)G(t,x+i\dr,\delta)
  -\Bigl(x-i\frac{1}{\dr}\Bigr)G(t,x-i\dr,\delta)
  =\frac{2i}{\dr}\Bigl(1+\delta\,t\frac{\partial}{\partial t}\Bigr)
  G(t,x,\delta).
  \label{feq_rclA_G}
\end{equation}
Since the solution of (\ref{feq_rclA}) is unique (up to normalization), it
is sufficient to show that (\ref{G_ans}) satisfies this functional equation.
This can be easily verified by explicit calculation, in which the
following formula derived from (\ref{G_ans}) is useful,
\begin{equation}
  G(t,x\pm i\dr,\delta)=\frac{1\pm i\dr\,t}{1\mp i\dr\,t}\,G(t,x,\delta).
\end{equation}

The explicit form of the inner product 
$(f,g)=\int_{-\infty}^{\infty} f(x)g(x){\mathbf w}(x,\delta)dx$ , 
{\em ie} the weight function ${\mathbf w}(x,\delta)$ is determined 
by using the generating function.
Here we list main results only without derivation.
Let us fix its normalisation by $(1,1)_{\delta}=\sqrt{\pi}$.
{}From the general theory (\ref{abh}), the orthogonality of
$H_n(x,\delta)$ is
\begin{equation}
  (H_n(x,\delta),H_m(x,\delta))_{\delta}=\delta_{n,m}h_n,\quad
  h_n=\sqrt{\pi}\,2^nn!
  \prod_{j=0}^{n-1}(1+\sfrac12 j\delta),
\end{equation}
which leads to
\begin{equation}
  (G(t,x,\delta),G(s,x,\delta))_{\delta}
  =\sqrt{\pi}\,(1-\delta\,ts)^{-\frac{2}{\delta}}.
  \label{(G,G)}
\end{equation}
Here we have used the identity
\begin{equation}
  F(x)=\sum_{n=0}^{\infty}\frac{x^n}{n!}\prod_{j=0}^{n-1}(1+\sfrac12 j\delta)
  =(1-\sfrac12\delta x)^{-\frac{2}{\delta}}.
\end{equation}
The weight
function is expressed as
\begin{eqnarray}
  {\mathbf w}(x,\delta)&\!\!=\!\!&
  \int_{-\infty}^{\infty}\frac{dt}{\sqrt{\pi}}
  \frac{\cos 2xt}{(\cosh\dr\,t)^{\frac{2}{\delta}}}
  =\frac{2^{\frac{2}{\delta}-1}}{\sqrt{\pi\delta}}
  B\Bigl(\frac{1}{\delta}+i\frac{x}{\dr},
  \frac{1}{\delta}-i\frac{x}{\dr}\Bigr)
  \label{wint}\\[8pt]
  &\!\!=\!\!&
  \frac{2^{\frac{2}{\delta}-1}}{\sqrt{\pi\delta}}
  \frac{\Gamma(\frac{1}{\delta}+i\frac{x}{\dr})
        \Gamma(\frac{1}{\delta}-i\frac{x}{\dr})}{\Gamma(\frac{2}{\delta})}
  =\frac{2^{\frac{2}{\delta}-1}}{\sqrt{\pi\delta}}
  \frac{|\Gamma(\frac{1}{\delta}+i\frac{x}{\dr})|^2}
       {\Gamma(\frac{2}{\delta})}.
\end{eqnarray}
The undeformed limit of the weight function 
$\lim_{\delta\to0}{\mathbf w}(x,\delta)=e^{-x^2}$ can be verified by
using the asymptotic expansion of the $\Gamma$-function.
The Taylor series of ${\mathbf w}(x,\delta)$ in powers of $\delta$ reads
\begin{eqnarray}
  {\mathbf w}(x,\delta)&\!\!=\!\!&e^{-x^2}\Bigl(
  1+\frac{\delta}{24}(3-12x^2+4x^4)
  +\frac{\delta^2}{5760}(45-1320x^2+2280x^4-864x^6+80x^8)\n
  &&\qquad
  +\frac{\delta^3}{2903040}(-14175-71820x^2+865620x^4-1042272x^6+386928x^8\n
  &&\qquad\qquad\qquad\quad
   -52416x^{10}+2240x^{12})+\cdots\Bigr).
\end{eqnarray}

Among many interesting properties of $H_n(x,\delta)$, we present only
\begin{eqnarray}
  \mbox{(i)}&&H_{2n}(0,\delta)=(-1)^n\,\frac{(2n)!}{n!}\,
  \prod_{j=0}^{n-1}(1+j\delta),\qquad\quad \mbox{see\ (\ref{originval})},\\
  \mbox{(ii)}&&\frac{d}{dx}H_n(x,\delta)=2\sum_{k=0}^{[\frac{n-1}{2}]}
  (2k)!{n\choose 2k+1}(-\delta)^kH_{n-1-2k}(x,\delta),
\end{eqnarray}
which is a deformation of ${d\over{dx}}H_n(x)=2nH_n(x)$.

Remark :
We may take the three-term recurrence (\ref{rec_rclA2}) as the definition
of the deformed Hermite polynomial $H_n(x,\delta)$ for an arbitrary
(complex) parameter $\delta$.
Then $H_n(x,\delta)$ is a polynomial in $x$ of degree $n$ and
in $\delta$ of degree $[\frac{n}{2}]$ with {\em integer\/} coefficients.
We will not repeat similar remarks which are valid for almost all of the
deformed polynomials in this paper.

%%%%%%%%%%%%%%%%%%%%%%%%%%%%
%  3.2                     %
%%%%%%%%%%%%%%%%%%%%%%%%%%%%
\subsection{Quadratic Confining Potential Case (2 Parameter Deformation)}
\label{sec:H.2}

For the solution $\{\bar{q}_j\}$ of the $A_{r-1}$ equation (\ref{rcqAeq}),
let us define
\begin{equation}
  \bar{q}_j=\sqrt{ag}\,y_j,\quad \delta=\frac{g}{a}\,,\quad
  \varepsilon=\frac{a}{b}\,,
  \label{para_rcqA}
\end{equation}
and introduce a degree $r$ polynomial in $x$ having zeros at
$\{y_j\}$:
\begin{equation}
  H_r(x,\delta,\varepsilon) \eqdef 2^r\prod_{j=1}^r(x-y_j).
\label{twoherm}
\end{equation}
It is a further deformation of the deformed Hermite polynomial defined
previously,
\begin{equation}
  \lim_{\varepsilon\rightarrow 0}H_r(x,\delta,\varepsilon)=H_r(x,\delta),
\quad
H_r(-x,\delta,\varepsilon)=(-1)^rH_r(x,\delta,\varepsilon).
\end{equation}
The symmetry between the two parameters $a\leftrightarrow b$ is
expressed as
\begin{equation}
  H_r(x,\delta\varepsilon,\varepsilon^{-1})=
  \varepsilon^{\frac{r}{2}}H_r(\varepsilon^{-\frac12}x,\delta,\varepsilon).
\end{equation}
If we define
$\hat{H}_r(x,\delta_1\eqdef\frac{g}{a}=\delta,
\delta_2\eqdef\frac{g}{b}=\frac{\delta}{\varepsilon})
\eqdef\sqrt{1+\varepsilon}^{\,r}
H_r(\frac{x}{\sqrt{1+\varepsilon}},\delta,\varepsilon)$,
then this symmetry is more manifest,
$\hat{H}_r(x,\delta_1,\delta_2)=\hat{H}_r(x,\delta_2,\delta_1)$.

The equation for the equilibrium (\ref{rcqAeq}) can be written as (we
replace $r$ by $n$)
\begin{equation}
  \prod_{k=1}^n
  {y_j-y_k-i\dr\over{y_j-y_k+i\dr}}=-
  {y_j+i\frac{1}{\dr}\over{y_j-i\frac{1}{\dr}}}\,
  {y_j+i\frac{1}{\varepsilon\dr}\over{y_j-i\frac{1}{\varepsilon\dr}}}\,.
  \label{rcqAeq2}
\end{equation}
{}From this equation, we obtain the following functional equation
for $H_n(x,\delta,\varepsilon)$ ($n\geq 1$),
\begin{eqnarray}
  &&\Bigl(x+i\frac{1}{\dr}\Bigr)\Bigl(x+i\frac{1}{\varepsilon\dr}\Bigr)
  H_n(x+i\dr,\delta,\varepsilon)
  +\Bigl(x-i\frac{1}{\dr}\Bigr)\Bigl(x-i\frac{1}{\varepsilon\dr}\Bigr)
  H_n(x-i\dr,\delta,\varepsilon)\n
  &&\quad=2(A_nx^2+B_nx+C_n)H_n(x,\delta,\varepsilon),
  \label{feq_rcqA}
\end{eqnarray}
because the LHS is a degree $n+2$ polynomial in $x$
with real coefficients which vanishes at $x=y_j$.
Here $A_n=A_n(\delta,\varepsilon)$, $B_n=B_n(\delta,\varepsilon)$ and
$C_n=C_n(\delta,\varepsilon)$ are real constants:
\begin{equation}
  A_n=1,\quad
  B_n=0,\quad
  C_n=-\Bigl((\frac{1}{\delta}+n)\varepsilon^{-1}
  +n+\sfrac12n(n-1)\delta\Bigr)\,.
  \label{feq_rcqA_coef}
\end{equation}
This functional equation contains all the information of the equilibrium.
The above functional equation (\ref{feq_rcqA}) reduces to
that of $H_n(x,\delta)$ (\ref{feq_rclA}) in a proper limit
$\varepsilon\to 0$.
This functional equation can be written as a difference equation,
\begin{equation}
  (1-\delta\,\varepsilon x^2)D_{x,\frac12i\dr}^2\,H_n(x,\delta,\varepsilon)
  -2(1+\varepsilon)xD_{x,i\dr}\,H_n(x,\delta,\varepsilon)
  +2n(1+(1+\sfrac{n-1}{2}\delta)\varepsilon)H_n(x,\delta,\varepsilon)=0\,.
\end{equation}

The functional equation (\ref{feq_rcqA}) implies (see Appendix)
the three-term recurrence (\ref{3termrec}) for the deformed Hermite 
polynomial $H_n(x,\delta,\varepsilon)$ with
\begin{eqnarray}
  &&a_n=0,\quad
  b_n=\frac{n}{2}\Bigl(1+\frac{n-1}{2}\delta\Bigr)
  \Bigl(1+\frac{n-1}{2}\delta\varepsilon\Bigr)
  \frac{d_{n}}{d_{2n-1}d_{2n+1}},\quad c_n=2^n,
  \label{rec_rcqA}\\
  &&d_m=1+\Bigl(1+\frac{m-2}{2}\delta\Bigr)\varepsilon,\\
  &&H_{n+1}(x,\delta,\varepsilon)-2xH_n(x,\delta,\varepsilon)
  +\Bigl(2n+n(n-1)\delta\Bigr)
  \frac{(1+\frac{n-1}{2}\delta\varepsilon)d_{n}}{d_{2n-1}d_{2n+1}}
  H_{n-1}(x,\delta,\varepsilon)=0.\qquad
  \label{rec_rcqA2}
\end{eqnarray}
Since $\delta$ and $\varepsilon$ are positive in this case
(\ref{para_rcqA}),
$b_n$ is also positive.
{}From this three-term recurrence we obtain the differential equation
for the generating function $\displaystyle G(t,x,\delta,\varepsilon)
=\sum_{n=0}^{\infty}\frac{t^n}{n!}H_n(x,\delta,\varepsilon)$,
\begin{equation}
  \Bigl((d_{2n-1}d_{2n+1})\Bigm|_{n\to t\frac{\partial}{\partial t}}
  (\frac{\partial}{\partial t}-2x)
  +(\frac{4}{n}\,d_{2n-1}d_{2n+1}b_n)
  \Bigm|_{n\to t\frac{\partial}{\partial t}}t\,\Bigr)
  G(t,x,\delta,\varepsilon)=0\,,
\end{equation}
which is a third order linear differential equation with respect to $t$.
The special case $\delta=0$  gives the original Hermite polynomial,
$H_n(x,0,\varepsilon)=(\sqrt{1+\varepsilon}\,)^{-n}
H_n(\sqrt{1+\varepsilon}\,x)$.
The value at the origin of the even polynomial shows a characteristic
deformation pattern, see (\ref{originval}):
\begin{equation}
  H_{2n}(0,\delta,\varepsilon)=(-1)^n\,\frac{(2n)!}{n!}\,
  \prod_{j=0}^{n-1}\frac{(1+j\delta)(1+j\delta\varepsilon)}
  {1+(1+(j+n-\frac12)\delta)\varepsilon}.
\end{equation}

%%%%%%%%%%%%%%%%%%%%%%%%%%%%%%%%%%%%%%%%%%%%%%%%%%%%%%%%%%%%%%%
%                                                             %
%  4. Deformations of the Laguerre Polynomial                 %
%                                                             %
%%%%%%%%%%%%%%%%%%%%%%%%%%%%%%%%%%%%%%%%%%%%%%%%%%%%%%%%%%%%%%%
\section{Deformation of the Laguerre Polynomial}
\label{sec:L}
\setcounter{equation}{0}

%%%%%%%%%%%%%%%%%%%%%%%%%%%%
%  4.1                     %
%%%%%%%%%%%%%%%%%%%%%%%%%%%%
\subsection{Linear Confining Potential Case (1 or 2 Parameter Deformation)}
\label{sec:L.1}

%%%%%%%%%%%%%
%  BCt_r    %
%%%%%%%%%%%%%
\paragraph{$\widetilde{BC}_r$ :}
For the solution $\{\bar{q}_j\}$ of the $\widetilde{BC}_r$ equation
(\ref{rclBCteq}), let us define
\begin{equation}
  \bar{q}_j=\sqrt{ag_0}\,y_j,\quad \delta=\frac{g_0}{a},\quad
  \alpha=\frac{g_1+g_2}{g_0}-1,\quad \gamma=\frac{g_1g_2}{g_0^{\,2}}\,,
  \label{para_rclBCt}
\end{equation}
and introduce a degree $r$ polynomial in $x$, having zeros at
$\{y_j^2\}$:
\begin{equation}
  L^{(\alpha)}_r(x,\gamma,\delta)\eqdef
  \frac{(-1)^r}{r!}\prod_{j=1}^r(x-y_j^2).
\label{twodefLag}
\end{equation}
It is a two parameter deformation of the Laguerre polynomial such that
\begin{equation}
  \lim_{\delta\to 0}L^{(\alpha)}_r(x,\gamma,\delta)=L^{(\alpha)}_r(x).
\end{equation}

The equation for the equilibrium (\ref{rclBCteq}) can be written as 
(we replace $r$ by $n$)
\begin{equation}
  \prod_{k=1}^n
  {(y_j-i\dr)^2-y_k^2\over{(y_j+i\dr)^2-y_k^2}}=
  {y_j-i\frac{\dr}{2}\over{y_j+i\frac{\dr}{2}}}\,
  {y_j+i\frac{1}{\dr}\over{y_j-i\frac{1}{\dr}}}\,
  {y_j^2+i(\alpha+1)\dr\,y_j-\gamma\delta\over
    {y_j^2-i(\alpha+1)\dr\,y_j-\gamma\delta}}\,.
  \label{rclBCteq2}
\end{equation}
{}From this equation, we obtain the following functional equation
for $L^{(\alpha)}_n(x,\gamma,\delta)$ ($n\geq 1$),
\begin{eqnarray}
  &&\frac{1}{y}\Biggl(
  \Bigl(y-i\frac{\dr}{2}\Bigr)\Bigl(y+i\frac{1}{\dr}\Bigr)
  \Bigl(y^2+i(\alpha+1)\dr\,y-\gamma\delta\Bigr)
  L^{(\alpha)}_n\Bigl((y+i\sqrt{\delta}\,)^2,\gamma,\delta\Bigr)\n
  &&\quad -
  \Bigl(y+i\frac{\dr}{2}\Bigr)\Bigl(y-i\frac{1}{\dr}\Bigr)
  \Bigl(y^2-i(\alpha+1)\dr\,y-\gamma\delta\Bigr)
  L^{(\alpha)}_n\Bigl((y-i\sqrt{\delta}\,)^2,\gamma,\delta\Bigr)
  \Biggr)\n
  &&=2i(A_ny^2+B_n)L^{(\alpha)}_n(y^2,\gamma,\delta),
  \label{feq_rclBCt}
\end{eqnarray}
because the LHS is $i$ times a degree $2n+2$ even polynomial in $y$
with real coefficients which vanishes at $y=\pm y_j$.
Here $A_n=A_n^{(\alpha)}(\gamma,\delta)$ and
$B_n=B^{(\alpha)}_n(\gamma,\delta)$ are real constants:
\begin{equation}
  A_n=\frac{1}{\sqrt{\delta}}
  \Bigl(1+(2n+\alpha+\sfrac12)\delta\Bigr),\quad
  B_n=\frac{\sqrt{\delta}}{2}\Bigl(
  \alpha+1-2\gamma+(n+\gamma)\delta\Bigr).
  \label{feq_rclBCt_coef}
\end{equation}
This functional equation contains all the information of the
equilibrium.
The functional equation can be written as a difference equation,
\begin{eqnarray}
  \hspace*{-5mm}&&\Bigl((1+(\alpha+\sfrac12)\delta)y^2
  +(\sfrac{\alpha+1}{2}-\gamma)\delta
  +\sfrac12\gamma\delta^2\Bigr)
  D_{y,\frac12i\dr}^2\,L^{(\alpha)}_n(y^2,\gamma,\delta)\n
  \hspace*{-5mm}&&
  +\Bigl(-2y^3+(2\alpha+1+(2\gamma-\alpha-1)\delta\,)y
  +\gamma\delta\,y^{-1}\Bigr)
  D_{y,i\dr}\,L^{(\alpha)}_n(y^2,\gamma,\delta)\n
  \hspace*{-5mm}&&
  +(4ny^2+n\delta)L^{(\alpha)}_n(y^2,\gamma,\delta)=0\,.
\end{eqnarray}
In the $\delta\to 0$ limit, it becomes,
\begin{equation}
  y\frac{d^2}{dy^2}L_n^{(\alpha)}(y^2)
  +(2\alpha+1-2y^2)\frac{d}{dy}L_n^{(\alpha)}(y^2)
  +4nyL_n^{(\alpha)}(y^2)=0,
\end{equation}
which is equivalent to the differential equation of the Laguerre
polynomial (\ref{diffeqL}).

By the same argument as $H_n(x,\delta)$, see Appendix,
the functional equation (\ref{feq_rclBCt}) implies
that the deformed Laguerre polynomial $L^{(\alpha)}_n(x,\gamma,\delta)$
satisfies the three-term recurrence (\ref{3termrec}) with
\begin{eqnarray}
  &&a_n=2n+\alpha+1
  +\Bigl(n(2n+1)+2n\alpha+\gamma\Bigr)\delta,
  \label{rec_rclBCt0}\\
  &&b_n=n(n+\alpha)
  \Bigl(1+(2n+\alpha-1)\delta+((n-1)(n+\alpha)+\gamma)\delta^2\Bigr),
  \ \ c_n={(-1)^n}/{n!},
  \label{rec_rclBCt}\\
  &&(n+1)L_{n+1}^{(\alpha)}(x,\gamma,\delta)
  +\Bigl(x-(2n+\alpha+1+(n(2n+1)+2n\alpha+\gamma)\delta)
  \Bigr)L_n^{(\alpha)}(x,\gamma,\delta)\n
  &&\qquad +(n+\alpha)
  \Bigl(1+(2n+\alpha-1)\delta+((n-1)(n+\alpha)+\gamma)\delta^2\Bigr)
  L_{n-1}^{(\alpha)}(x,\gamma,\delta)=0\,.
  \label{rec_rclBCt2}
\end{eqnarray}
In this case (\ref{para_rclBCt}), the parameter ranges  are
$\delta,\gamma>0$ and $\alpha>-1$. So $b_n$ is positive.
Therefore $L^{(\alpha)}_n(x,\gamma,\delta)$ is a set of orthogonal
polynomials with respect to some positive definite inner product.
{}From this three-term recurrence we obtain the differential equation
for the generating function
$\displaystyle G^{(\alpha)}(t,x,\gamma,\delta)
=\sum_{n=0}^{\infty}t^nL^{(\alpha)}_n(x,\gamma,\delta)$,
\begin{equation}
  \Bigl(\frac{\partial}{\partial t}
  +x-a_n\Bigm|_{n\to t\frac{\partial}{\partial t}}
  +\frac{b_n}{n}\Bigm|_{n\to t\frac{\partial}{\partial t}}t\,\Bigr)
  G^{(\alpha)}(t,x,\gamma,\delta)=0\,,
\end{equation}
which is a third order linear differential equation with respect to $t$.

%%%%%%%%%%%%%
%  C_r      %
%%%%%%%%%%%%%
\paragraph{$C_r$ :}
For the solution $\{\bar{q}_j\}$ of the $C_r$ equation (\ref{rclCeq}),
let us define
\begin{equation}
  \bar{q}_j=\sqrt{ag_S}\,y_j\,,\quad
  \delta=\frac{g_S}{a}\,,\quad \alpha=\frac{g_L}{g_S}-1\,,
  \label{para_rclC}
\end{equation}
and introduce a degree $r$ polynomial in $x$, having zeros at
$\{y_j^2\}$:
\begin{equation}
  L^{(\alpha)}_r(x,\delta)\eqdef\frac{(-1)^r}{r!}\prod_{j=1}^r(x-y_j^2).
  \label{l_al_x_del}
\end{equation}
This is a deformation of the Laguerre polynomial such that
$\displaystyle\lim_{\delta\to 0}L^{(\alpha)}_r(x,\delta)=L^{(\alpha)}_r(x)$,
and obviously it is a special case of $L^{(\alpha)}_n(x,\gamma,\delta)$,
\begin{equation}
  L^{(\alpha)}_n(x,\delta)=L^{(\alpha)}_n(x,0,\delta).
\end{equation}

The functional equation for $L^{(\alpha)}_n(x,\delta)$ is easily 
obtained from that of $L^{(\alpha)}_n(x,\gamma,\delta)$ 
(\ref{feq_rclBCt}) and will not be presented.
The three-term recurrence  for $L^{(\alpha)}_n(x,\delta)$ reads
\begin{eqnarray}
  &&(n+1)L_{n+1}^{(\alpha)}(x,\delta)
  +\Bigl(x-(2n+\alpha+1+n(2n+2\alpha+1)\delta)\Bigr)
  L_n^{(\alpha)}(x,\delta)\nonumber\\
  &&\qquad\qquad +(n+\alpha)\Bigl(1+(n-1)\delta\Bigr)
  \Bigl(1+(n+\alpha)\delta\Bigr)
  L_{n-1}^{(\alpha)}(x,\delta)=0\,,
  \label{rec_rclC2}
\end{eqnarray}

The value at the origin shows a simple deformation pattern
\begin{equation}
  L_n^{(\alpha)}(0,\delta)={n+\alpha\choose n}
  \prod_{j=0}^{n-1}(1+j\delta).
\end{equation}

%%%%%%%%%%%%%
%  B_r      %
%%%%%%%%%%%%%
\paragraph{$B_r$ :}
For the solution $\{\bar{q}_j\}$ of the $B_r$ equation (\ref{rclBeq}),
let us define
\begin{equation}
  \bar{q}_j=\sqrt{ag_L}\,y_j\,,\quad \delta=\frac{g_L}{a}\,,\quad
  \alpha=\frac{g_S}{g_L}-1\,,
  \label{para_rclB}
\end{equation}
and introduce a degree $r$ polynomial in $x$, having zeros at
$\{y_j^2\}$:
\begin{equation}
  \tilde{L}^{(\alpha)}_r(x,\delta)
  \eqdef\frac{(-1)^r}{r!}\prod_{j=1}^r(x-y_j^2).
  \label{ltilda_al_x_del}
\end{equation}
This is a deformation of the Laguerre polynomial such that
$\displaystyle\lim_{\delta\to 0}\tilde{L}^{(\alpha)}_r(x,0)
=L^{(\alpha)}_r(x)$, and 
obviously it is a special case of $L^{(\alpha)}_n(x,\gamma,\delta)$
\begin{equation}
  \tilde{L}^{(\alpha)}_n(x,\delta)
  =L^{(\alpha)}_n(x,\sfrac14(\alpha+1)^2,\delta).
\end{equation}
The functional equation of $\tilde{L}^{(\alpha)}_n(x,\delta)$ is 
easily obtained from that of $L^{(\alpha)}_n(x,\gamma,\delta)$ 
(\ref{feq_rclBCt}) and will not be presented.
The three-term recurrence for $\tilde{L}^{(\alpha)}_n(x,\delta)$ reads
\begin{eqnarray}
  &&(n+1)\tilde{L}_{n+1}^{(\alpha)}(x,\delta)
  +\Bigl(x-(2n+\alpha+1+(n(2n+2\alpha+1)+\sfrac14(\alpha+1)^2)\delta)\Bigr)
  \tilde{L}_n^{(\alpha)}(x,\delta)\nonumber\\
  &&\qquad\qquad\qquad\qquad
  +(n+\alpha)\Bigl(1+(n+\sfrac{\alpha-1}{2})\delta\Bigr)^{\!2}
  \tilde{L}_{n-1}^{(\alpha)}(x,\delta)=0.
  \label{rec_rclB2}
\end{eqnarray}

\medskip
%%%%%%%%%%%%%
%  D_r      %
%%%%%%%%%%%%%
\paragraph{$D_r$ :}
As in the Calogero systems, the $D_r$ is a special case $g_S=0$ of
the $B_r$ theory described by
$\tilde{L}_r^{(-1)}(x,\delta)=L_r^{(-1)}(x,\delta)$,
which has a zero at $x=0$ for all $r$.

%%%%%%%%%%%%%%%%%%%%%%%%%%%%
%  4.2                     %
%%%%%%%%%%%%%%%%%%%%%%%%%%%%
\subsection{Quadratic Confining Potential Case (2 or 3 Parameter Deformation)}
\label{sec:L.2}

%%%%%%%%%%%%%
%  BCt_r    %
%%%%%%%%%%%%%
\paragraph{$\widetilde{BC}_r$:}

For the solution $\{\bar{q}_j\}$ of the $\widetilde{BC}_r$ equation
(\ref{rcqBCteq}), let us define
\begin{equation}
  \bar{q}_j=\sqrt{ag_0}\,y_j,\quad
  \delta=\frac{g_0}{a},\quad \varepsilon=\frac{b}{a},\quad
  \alpha=\frac{g_1+g_2}{g_0}-1,\quad \gamma=\frac{g_1g_2}{g_0^{\,2}}\,,
  \label{para_rcqBCt}
\end{equation}
and introduce a degree $r$ polynomial in $x$, having zeros at
$\{y_j^2\}$:
\begin{equation}
  L^{(\alpha)}_r(x,\gamma,\delta,\varepsilon)
  \eqdef\frac{(-1)^r}{r!}\prod_{j=1}^r(x-y_j^2).
\label{threedefLag}
\end{equation}
It is a further deformation of the deformed Laguerre polynomial defined
previously,
\begin{equation}
  \lim_{\varepsilon\rightarrow 0}
  L^{(\alpha)}_r(x,\gamma,\delta,\varepsilon)
  =L^{(\alpha)}_r(x,\gamma,\delta).
\end{equation}
The symmetry between the two parameters $a\leftrightarrow b$ is
expressed as
\begin{equation}
  L^{(\alpha)}_r(x,\gamma,\delta\varepsilon,\varepsilon^{-1})=
  \varepsilon^{r}
  L^{(\alpha)}_r(\varepsilon^{-1}x,\gamma,\delta,\varepsilon).
\end{equation}
If we define
$\hat{L}^{(\alpha)}_r
(x,\gamma,\delta_1\eqdef\frac{g}{a}=\delta,
\delta_2\eqdef\frac{g}{b}=\frac{\delta}{\varepsilon})
\eqdef(1+\varepsilon)^r
L^{(\alpha)}_r(\frac{x}{1+\varepsilon},\gamma,\delta,\varepsilon)$,
then this symmetry is more manifest,
$\hat{L}^{(\alpha)}_r(x,\gamma,\delta_1,\delta_2)
=\hat{L}^{(\alpha)}_r(x,\gamma,\delta_2,\delta_1)$.

The equation for the equilibrium (\ref{rcqBCteq}) can be written as (we
replace $r$ by $n$)
\begin{equation}
  \prod_{k=1}^n
  {(y_j-i\dr)^2-y_k^2\over{(y_j+i\dr)^2-y_k^2}}=-
  {y_j-i\frac{\dr}{2}\over{y_j+i\frac{\dr}{2}}}\,
  {y_j+i\frac{1}{\dr}\over{y_j-i\frac{1}{\dr}}}\,
  {y_j+i\frac{1}{\varepsilon\dr}\over{y_j-i\frac{1}{\varepsilon\dr}}}\,
  {y_j^2+i(\alpha+1)\dr\,y_j-\gamma\delta\over
    {y_j^2-i(\alpha+1)\dr\,y_j-\gamma\delta}}\,.
  \label{rcqBCteq2}
\end{equation}
The equivalent functional equation
for $L^{(\alpha)}_n(x,\gamma,\delta,\varepsilon)$ ($n\geq 1$) reads
\begin{eqnarray}
  \hspace*{-10mm}&&\frac{1}{y}\Biggl(
  \Bigl(y-i\frac{\dr}{2}\Bigr)
  \Bigl(y+i\frac{1}{\dr}\Bigr)\Bigl(y+i\frac{1}{\varepsilon\dr}\Bigr)
  \Bigl(y^2+i(\alpha+1)\dr\,y-\gamma\delta\Bigr)
  L^{(\alpha)}_n\Bigl((y+i\sqrt{\delta}\,)^2,\gamma,\delta,\varepsilon\Bigr)\n
  \hspace*{-10mm}&&\quad +
  \Bigl(y+i\frac{\dr}{2}\Bigr)
  \Bigl(y-i\frac{1}{\dr}\Bigr)\Bigl(y-i\frac{1}{\varepsilon\dr}\Bigr)
  \Bigl(y^2-i(\alpha+1)\dr\,y-\gamma\delta\Bigr)
  L^{(\alpha)}_n\Bigl((y-i\sqrt{\delta}\,)^2,\gamma,\delta,\varepsilon\Bigr)
  \Biggr)\n
  \hspace*{-10mm}&&=2(A_ny^4+B_ny^2+C_n)
  L^{(\alpha)}_n(y^2,\gamma,\delta,\varepsilon),
  \label{feq_rcqBCt}
\end{eqnarray}
because the LHS is a degree $2n+4$ even polynomial in $y$
with real coefficients which vanishes at $y=\pm y_j$.
Here  $A_n=A^{(\alpha)}_n(\gamma,\delta,\varepsilon)$,
$B_n=B^{(\alpha)}_n(\gamma,\delta,\varepsilon)$ and
$C_n=C^{(\alpha)}_n(\gamma,\delta,\varepsilon)$ are real constants:
\begin{eqnarray}
  A_n&\!\!=\!\!&1,\\
  B_n&\!\!=\!\!&
  -(\delta^{-1}+2n+\alpha+\sfrac12)\varepsilon^{-1}
  -(2n+\alpha+\sfrac12)
  -\Bigl(\gamma-\sfrac{\alpha+1}{2}+2n(n+\alpha)\Bigr)\delta,\\
  C_n&\!\!=\!\!&
  -\frac12\Bigl(\alpha+1-2\gamma
  +(n+\gamma)\delta\Bigr)\varepsilon^{-1}
  -\frac12(n+\gamma)\delta-\frac12n(n+\alpha)\delta^2.
  \label{feq_rcqBCt_coef}
\end{eqnarray}
This functional equation contains all the information of the
equilibrium. In a proper limit
$\varepsilon\to 0$, the above functional equation (\ref{feq_rcqBCt})
reduces to that of $L^{(\alpha)}_n(x,\gamma,\delta)$ (\ref{feq_rclBCt}).
This functional equation can be written as a difference equation  as
previous examples.

The functional equation (\ref{feq_rcqBCt}) implies  (see Appendix)
that the deformed Laguerre polynomial
$L^{(\alpha)}_n(x,\gamma,\delta,\varepsilon)$
satisfies the three-term recurrence (\ref{3termrec}) with,
\begin{eqnarray}
  a_n&\!\!=\!\!&
  2n+\alpha+1+\frac{X_0+X_1\varepsilon+X_2\varepsilon^2}
  {d_{2n}d_{2n+2}}\,,
\label{rec_rcqBCt0}\\
  b_n&\!\!=\!\!&
  n(n+\alpha)
  \Bigl(1+(2n+\alpha-1)\delta+((n-1)(n+\alpha)+\gamma)\delta^2\Bigr)\n
  &&\times
  \Bigl(1+(1+(n-1)\delta)\varepsilon\Bigr)
  \Bigl(1+(2n+\alpha-1)\delta\varepsilon
  +((n-1)(n+\alpha)+\gamma)\delta^2\varepsilon^2\Bigr)\n
  &&\times
  \frac{d_n}{d_{2n-1}d_{2n}^{\ 2}d_{2n+1}},\quad c_n={(-1)^n}/{n!},
  \label{rec_rcqBCt}\\
  d_m&\!\!=\!\!&1+\Bigl(1+(m+\alpha-1)\delta\Bigr)\varepsilon.
\end{eqnarray}
Here
$X_0$, $X_1$ and $X_2$ are
\begin{eqnarray}
  X_0&\!\!=\!\!&
  \Bigl(n(2n+1)+2n\alpha+\gamma\Bigr)\delta,\\
  X_1&\!\!=\!\!&
  -(2n+\alpha+1)
  -\Bigl(2n(n+\alpha+1)+(\alpha+1)^2-2\gamma\Bigr)\delta\n
  &&+\Bigl(n(4n^2+(6\alpha+1)n+2\alpha^2-1)
  +(2n+\alpha-1)\gamma\Bigr)\delta^2,\\
  X_2&\!\!=\!\!&
  -(2n+\alpha+1)-\Bigl(6n^2+3(2\alpha+1)n+2\alpha(\alpha+1)-\gamma
  \Bigr)\delta \n
  &&-\Bigl(4n^3+3(2\alpha+1)n^2+(4\alpha^2+4\alpha-1)n
  +(\alpha-1)(\alpha+1)^2-(2n+\alpha-1)\gamma\Bigr)\delta^2 \n
  &&+n(n+\alpha)\Bigl(2n(n+\alpha)-\alpha-1+2\gamma\Bigr)\delta^3\,.
\end{eqnarray}
Namely $L^{(\alpha)}_n(x,\gamma,\delta,\varepsilon)$ satisfies
\begin{equation}
  (n+1)L_{n+1}^{(\alpha)}(x,\gamma,\delta,\varepsilon)
  +(x-a_n)L_n^{(\alpha)}(x,\gamma,\delta,\varepsilon)
  +\frac{1}{n}\,b_n\,L_{n-1}^{(\alpha)}(x,\gamma,\delta,\varepsilon)=0.
  \label{rec_rcqBCt2}
\end{equation}
In this case (\ref{para_rcqBCt}), the parameter ranges are 
$\delta,\varepsilon,\gamma>0$ and $\alpha>-1$. So $b_n$ is positive.
{}From this three-term recurrence we obtain the differential equation
for the generating function
$\displaystyle G^{(\alpha)}(t,x,\gamma,\delta,\varepsilon)
=\sum_{n=0}^{\infty}t^nL^{(\alpha)}_n(x,\gamma,\delta,\varepsilon)$,
\begin{eqnarray}
  \hspace*{-10mm}&&\Bigl((d_{2n-1}d_{2n}^{\ 2}d_{2n+1}d_{2n+2})
  \Bigm|_{n\to t\frac{\partial}{\partial t}}
  (\frac{\partial}{\partial t}+x)
  -(d_{2n-1}d_{2n}^{\ 2}d_{2n+1}d_{2n+2}a_n)
  \Bigm|_{n\to t\frac{\partial}{\partial t}}\n
  \hspace*{-10mm}&&\qquad\qquad\qquad\qquad
  +(\frac{1}{n}\,d_{2n-1}d_{2n}^{\ 2}d_{2n+1}d_{2n+2}b_n)
  \Bigm|_{n\to t\frac{\partial}{\partial t}}t\,\Bigr)
  G^{(\alpha)}(t,x,\gamma,\delta,\varepsilon)=0\,,
\end{eqnarray}
which is an 8-th order linear differential equation with respect to
$t$. The special case of $\delta=0$  gives the original Laguerre polynomial,
$L_n^{(\alpha)}(x,\gamma,0,\varepsilon)=(1+\varepsilon\,)^{-n}
L_n^{(\alpha)}((1+\varepsilon\,)x)$.

%%%%%%%%%%%%%
%  C_r      %
%%%%%%%%%%%%%
\paragraph{$C_r$ :}
For the solution $\{\bar{q}_j\}$ of the $C_r$ equation (\ref{rcqCeq}),
let us define
\begin{equation}
  \bar{q}_j=\sqrt{ag_S}\,y_j,\quad \delta=\frac{g_S}{a},\quad
  \varepsilon=\frac{b}{a},\quad \alpha=\frac{g_L}{g_S}-1,
  \label{para_rcqC}
\end{equation}
and introduce a degree $r$ polynomial in $x$, having zeros at
$\{y_j^2\}$ :
\begin{equation}
  L_r^{(\alpha)}(x,\delta,\varepsilon)
  \eqdef\frac{(-1)^r}{r!}\prod_{j=1}^r(x-y_j^2).
\end{equation}
This is a further deformation of the deformed Laguerre polynomial defined
previously such that $\displaystyle\lim_{\varepsilon\to 0}
L_r^{(\alpha)}(x,\delta,\varepsilon)
=L_r^{(\alpha)}(x,\delta)$ (\ref{l_al_x_del}),
and obviously it is a special case of
$L_r^{(\alpha)}(x,\gamma,\delta,\varepsilon)$,
\begin{equation}
  L_r^{(\alpha)}(x,\delta,\varepsilon)
  =L_r^{(\alpha)}(x,0,\delta,\varepsilon).
\end{equation}
The value at the origin shows a characteristic deformation pattern
\begin{equation}
  L_n^{(\alpha)}(0,\delta,\varepsilon)={n+\alpha\choose n}
  \prod_{j=0}^{n-1}\frac{(1+j\delta)(1+j\delta\varepsilon)}
  {1+(1+(\alpha+j+r)\delta)\varepsilon}\,.
\end{equation}

%%%%%%%%%%%%%
%  B_r      %
%%%%%%%%%%%%%
\paragraph{$B_r$ :}
For the solution $\{\bar{q}_j\}$ of the $B_r$ equation (\ref{rcqBeq}),
let us define
\begin{equation}
  \bar{q}_j=\sqrt{ag_L}\,y_j,\quad \delta=\frac{g_L}{a},\quad
  \varepsilon=\frac{b}{a},\quad \alpha=\frac{g_S}{g_L}-1,
  \label{para_rcqB}
\end{equation}
and introduce a degree $r$ polynomial in $x$, having zeros at
$\{y_j^2\}$ :
\begin{equation}
  \tilde{L}_r^{(\alpha)}(x,\delta,\varepsilon)
  \eqdef\frac{(-1)^r}{r!}\prod_{j=1}^r(x-y_j^2).
\end{equation}
This is a further deformation of the deformed Laguerre polynomial defined
previously such that $\displaystyle\lim_{\varepsilon\to 0}
\tilde{L}_r^{(\alpha)}(x,\delta,\varepsilon)
=\tilde{L}_r^{(\alpha)}(x,\delta)$ (\ref{ltilda_al_x_del}),
and obviously it is a special case of
$L_r^{(\alpha)}(x,\gamma,\delta,\varepsilon)$,
\begin{equation}
  \tilde{L}_r^{(\alpha)}(x,\delta,\varepsilon)
  =L_r^{(\alpha)}(x,\sfrac14(\alpha+1)^2,\delta,\varepsilon).
\end{equation}

%%%%%%%%%%%%%
%  D_r      %
%%%%%%%%%%%%%
\paragraph{$D_r$ :}
As in the Calogero systems, the $D_r$ is a special case $g_S=0$ of the
$B_r$ theory described by
$\tilde{L}_r^{(-1)}(x,\delta,\varepsilon)=
{L}_r^{(-1)}(x,\delta,\varepsilon)$,
which has a zero at $x=0$ for all $r$.

\paragraph{Deformation of the identities :}
Before going to the systems with trigonometric potentials,
let us present the one and two parameter deformation of the 
{\em identities\/} between the Hermite and Laguerre polynomials,
(\ref{LHiden}) and (\ref{LHiden2}), which could be considered as
consequences of the folding (\ref{acfold})--(\ref{bcfold}) and
(\ref{AtoC})--(\ref{DtoB}) of the rational potentials. 
The one and two parameter deformation of the even degree identities
(\ref{LHiden}) are
\begin{eqnarray}
  2^{-2r}H_{2r}(x,\delta)&\!\!=\!\!&
  (-1)^r r!\, L_r^{(-\frac12)}(x^2,\delta),
  \label{LHidendef1}\\
  2^{-2r}H_{2r}(x,\delta,\varepsilon)&\!\!=\!\!&
  (-1)^r r!\, L_r^{(-\frac12)}(x^2,\delta,\varepsilon),
  \label{LHidendef2}
\end{eqnarray}
which are connected with the folding $A_{2r-1}\to C_r$ (\ref{acfold}).
The one and two parameter deformation of the odd degree identities
(\ref{LHiden2}) are
\begin{eqnarray}
  2^{-2r-1}H_{2r+1}(x,\delta)&\!\!=\!\!&
  x\,(-1)^r r!\, L_r^{(\frac12)}(x^2,\sfrac12,\delta),
  \label{LHiden2def1}\\
  2^{-2r-1}H_{2r+1}(x,\delta,\varepsilon)&\!\!=\!\!&
  x\,(-1)^r r!\, L_r^{(\frac12)}(x^2,\sfrac12,\delta,\varepsilon),
  \label{LHiden2def2}
\end{eqnarray}
which are related to the folding $A_{2r}\to \widetilde{BC}_r$
(\ref{bcfold}).
The one and two parameter deformation of the
identities between Laguerre polynomials (\ref{lagsiden}) are
\begin{eqnarray}
  (r+1)\,\tilde{L}^{(-1)}_{r+1}(x,\delta)&\!\!=\!\!&
  -x\,\tilde{L}^{(1)}_r(x,\delta),
  \label{deforLHiden}\\
  (r+1)\,\tilde{L}^{(-1)}_{r+1}(x,\delta,\varepsilon)&\!\!=\!\!&
  -x\,\tilde{L}^{(1)}_r(x,\delta,\varepsilon),
  \label{2deforLHiden}
\end{eqnarray}
which are related to the folding $D_{r+1}\to {B}_r$ (\ref{dbfold}).

%%%%%%%%%%%%%%%%%%%%%%%%%%%%%%%%%%%%%%%%%%%%%%%%%%%%%%%%%%%%%%%
%                                                             %
%  5. Deformations of the Jacobi Polynomial                   %
%                                                             %
%%%%%%%%%%%%%%%%%%%%%%%%%%%%%%%%%%%%%%%%%%%%%%%%%%%%%%%%%%%%%%%
\section{Deformation of the Jacobi Polynomial}
\label{sec:J}
\setcounter{equation}{0}

%%%%%%%%%%%%%%%%%%%%%%%%%%%%
%  5.1 A_{r-1}             %
%%%%%%%%%%%%%%%%%%%%%%%%%%%%
\subsection{$A_{r-1}$}
\label{nodeform}

For the $A_{r-1}$ case, the equilibrium position, {\em ie\/}
the solution $\{\bar{q}_j\}$ of the $A_{r-1}$ equation (\ref{rsAeq}),
is the same as the original Sutherland system (\ref{eqspaced1}).
Therefore the polynomial describing the equilibrium is same as the 
original Sutherland system, the Chebyshev polynomial of the first
kind $T_r(x)$ (\ref{archev}).
In other words the Chebyshev polynomials are not {\em deformed\/}
in the present scheme.

%%%%%%%%%%%%%%%%%%%%%%%%%%%%
%  5.2                     %
%%%%%%%%%%%%%%%%%%%%%%%%%%%%
\subsection{$B'C_r$}

%%%%%%%%%%%%%
%  BC_r     %
%%%%%%%%%%%%%
\paragraph{$B'C_r$ :}
For the solution $\{\bar{q}_j\}$ of the $B'C_r$ equation (\ref{rsBCeq}),
let us define
\begin{equation}
  \delta=\tanh^2\!g_M,\quad
  \alpha=\frac{\tanh g_S}{\tanh g_M}+\frac{\tanh 2g_L}{2\tanh g_M}-1,\quad
  \beta=\frac{\tanh 2g_L}{2\tanh g_M}-1,
  \label{para_rsBC}
\end{equation}
and introduce a degree $r$ polynomial in $x$ having zeros at
$\{\cos 2\bar{q}_j\}$ :
\begin{equation}
  P_r^{(\alpha,\beta)}(x,\delta)
  \eqdef 2^{-r}\schoose{\alpha+\beta+2r}{r}
  \prod_{j=1}^r(x-\cos 2\bar{q}_j)\,.
\label{defjacobidef}
\end{equation}
It is a deformation of the Jacobi polynomial (\ref{Pdef}) such that
\begin{equation}
  \lim_{\delta\to 0}P_r^{(\alpha,\beta)}(x,\delta)=P_r^{(\alpha,\beta)}(x).
\end{equation}

The equation for the equilibrium (\ref{rsBCeq}) can be written as 
(we replace $r$ by $n$)
\begin{eqnarray}
  &&\prod_{k=1}^n
  \frac{\frac{1+\delta}{1-\delta}\cos 2\bar{q}_j
        +i\,\frac{2\dr}{1-\delta}\sin 2\bar{q}_j-\cos 2\bar{q}_k}
       {\frac{1+\delta}{1-\delta}\cos 2\bar{q}_j
        -i\,\frac{2\dr}{1-\delta}\sin 2\bar{q}_j-\cos 2\bar{q}_k}\n
  &=\!\!&
  \frac{\dr\cos 2\bar{q}_j+i\sin 2\bar{q}_j}
       {\dr\cos 2\bar{q}_j-i\sin 2\bar{q}_j}\,
  \frac{\frac{\sin 2\bar{q}_j}{1+\cos 2\bar{q}_j}+i\,(\alpha-\beta)\dr\,}
       {\frac{\sin 2\bar{q}_j}{1+\cos 2\bar{q}_j}-i\,(\alpha-\beta)\dr\,}\,
  \frac{\frac{\sin 2\bar{q}_j}{\cos 2\bar{q}_j}+i\,2(\beta+1)\dr\,}
       {\frac{\sin 2\bar{q}_j}{\cos 2\bar{q}_j}-i\,2(\beta+1)\dr\,}\,.
  \label{rsBCeq2}
\end{eqnarray}
Since $\bar{q}_j$ can be restricted to $0\leq\bar{q}_j\leq{\pi}/{2}$,
$\sin 2\bar{q}_j$ is $\sin 2\bar{q}_j=\sqrt{1-\cos^22\bar{q}_j}$.
{}From this equation, for $-1\leq x\leq 1$,
we obtain the following functional equation
for $P^{(\alpha,\beta)}_n(x,\delta)$ ($n\geq 1$),
\begin{eqnarray}
  &&\phantom{-}\Bigl(\dr\,x-i\sqrt{1-x^2}\Bigr)
  \Bigl((\alpha-\beta)\dr\,(1+x)+i\sqrt{1-x^2}\Bigr)
  \Bigl(2(\beta+1)\dr\,x+i\sqrt{1-x^2}\Bigr)\n
  &&\qquad\times
  P_n^{(\alpha,\beta)}
  \Bigl(\sfrac{1+\delta}{1-\delta}\,x
  +i\,\sfrac{2\dr}{1-\delta}\sqrt{1-x^2},\delta\Bigr)\n
  &&-
  \Bigl(\dr\,x+i\sqrt{1-x^2}\Bigr)
  \Bigl((\alpha-\beta)\dr\,(1+x)-i\sqrt{1-x^2}\Bigr)
  \Bigl(2(\beta+1)\dr\,x-i\sqrt{1-x^2}\Bigr)\n
  &&\qquad\times
  P_n^{(\alpha,\beta)}
  \Bigl(\sfrac{1+\delta}{1-\delta}\,x
  -i\,\sfrac{2\dr}{1-\delta}\sqrt{1-x^2},\delta\Bigr)\n
  &&=
  2i\sqrt{1-x^2}\,
  (A_nx^2+B_nx+C_n)P_n^{(\alpha,\beta)}(x,\delta),
  \label{feq_rsBC}
\end{eqnarray}
because the LHS is $i\sqrt{1-x^2}$ times a degree $n+2$ polynomial in
$x$ with real coefficients which vanishes at $x=\cos 2\bar{q}_j$.
Here $A_n=A^{(\alpha,\beta)}_n(\delta)$,
$B_n=B^{(\alpha,\beta)}_n(\delta)$ and $C_n=C^{(\alpha,\beta)}_n(\delta)$
are real constants:
\begin{eqnarray}
  A_n&\!\!=\!\!&
  -(1-\delta\,)^{-(n-1)}\times
   \frac12\Bigl((1+(\alpha-\beta)\dr\,)(1+2(\beta+1)\dr\,)(1+\dr\,)^{2n-1}\n
  &&\qquad\qquad\qquad\qquad\quad
  +(1-(\alpha-\beta)\dr\,)(1-2(\beta+1)\dr\,)(1-\dr\,)^{2n-1}\Bigr),\qquad\\
  B_n&\!\!=\!\!&
  -(\alpha-\beta)(1+2\beta)\delta,\\
  C_n&\!\!=\!\!&
  (1-\delta\,)^{-n}\times
   \frac12\Bigl((1+(\alpha-\beta)\dr\,)(1+2(\beta+1)\dr\,)(1+\dr\,)^{2n-1}\n
  &&\qquad\qquad\qquad
  +(1-(\alpha-\beta)\dr\,)(1-2(\beta+1)\dr\,)(1-\dr\,)^{2n-1}\n
  &&\qquad\qquad\qquad
  -2(\alpha-\beta-1)(1+2\beta)\delta\,(1-\delta\,)^{n-1}\Bigr).
  \label{feq_rsBC_coef}
\end{eqnarray}
The functional equation (\ref{feq_rsBC}) contains all the information of
the equilibrium.
In the $\delta\to 0$ limit, this functional equation reduces to the
differential equation of the Jacobi polynomial (\ref{diffeqJ}).

The three-term recurrence (\ref{3termrec}) for the deformed Jacobi
polynomial $P^{(\alpha,\beta)}_n(x,\delta)$ is a consequence of the
functional equation (\ref{feq_rsBC}) (see Appendix).
The constants in (\ref{3termrec}) are
\begin{eqnarray}
  a_n&\!\!=\!\!&
  (\beta-\alpha)(1-\delta)^{n-1}\n
  &&\times
   \frac12\Bigl(
   (1+2\beta)\Bigl(
       (1+(\alpha-\beta)\dr\,)(1+2(\beta+1)\dr\,)(1+\dr\,)^{2n-1}\n
  &&\qquad\qquad\qquad
        +(1-(\alpha-\beta)\dr\,)(1-2(\beta+1)\dr\,)(1-\dr\,)^{2n-1}\Bigr)\n
  &&\qquad
   +2(\alpha-\beta-1)(1-4(\beta+1)^2\delta\,)(1-\delta\,)^{n-1}\Bigr)
   \times\frac{1}{d_{2n}\,d_{2n+2}}\,,
\label{rec_rsBCA}\\
  b_n&\!\!=\!\!&
  \frac{1}{2\dr}\Bigl((1+\dr\,)^n-(1-\dr\,)^n\Bigr) \n
  &&\times
  \frac12\Bigl((1+2(\beta+1)\dr\,)(1+\dr\,)^{n-1}
     +(1-2(\beta+1)\dr\,)(1-\dr\,)^{n-1}\Bigr) \n
  &&\times
  \frac{1}{2\dr}\Bigl((1+2(\beta+1)\dr\,)(1+\dr\,)^{2n-2}
     -(1-2(\beta+1)\dr\,)(1-\dr\,)^{2n-2}\Bigr) \n
  &&\times
  \frac12\Bigl((1+(\alpha-\beta)\dr\,)(1+\dr\,)^{n-1}
     +(1-(\alpha-\beta)\dr\,)(1-\dr\,)^{n-1}\Bigr) \n
  &&\times
  \frac{1}{2\dr}\Bigl((1+(\alpha-\beta)\dr\,)^2(1+2(\beta+1)\dr\,)
       (1+\dr\,)^{2n-2}\n
  &&\qquad\quad
     -(1-(\alpha-\beta)\dr\,)^2(1-2(\beta+1)\dr\,)(1-\dr\,)^{2n-2}\Bigr) \n
  &&\times
  \frac{d_{n}}{d_{2n-1}\,d_{2n}^{\ 2}\,d_{2n+1}},
  \quad c_n=2^{-n}\schoose{\alpha+\beta+2n}{n},
  \label{rec_rsBC}\\
  d_m&\!\!=\!\!&\frac{1}{2\dr}\Bigl(
  (1+(\alpha-\beta)\dr\,)(1+2(\beta+1)\dr\,)(1+\dr\,)^{m-2}\n
  &&\qquad
  -(1-(\alpha-\beta)\dr\,)(1-2(\beta+1)\dr\,)(1-\dr\,)^{m-2}\Bigr).
\end{eqnarray}
Namely $P^{(\alpha,\beta)}_n(x,\delta)$ satisfies
\begin{eqnarray}
  &&\frac{2(n+1)(\alpha+\beta+n+1)}{(\alpha+\beta+2n+1)(\alpha+\beta+2n+2)}
  P_{n+1}^{(\alpha,\beta)}(x,\delta)
  +(x-a_n)P_n^{(\alpha,\beta)}(x,\delta)\n
  &&\qquad\qquad
  +\frac{(\alpha+\beta+2n-1)(\alpha+\beta+2n)}{2n(\alpha+\beta+n)}\,
  b_n\,P_{n-1}^{(\alpha,\beta)}(x,\delta)=0.
  \label{rec_rsBC2}
\end{eqnarray}
In this case (\ref{para_rsBC}), the parameter ranges are
$\delta>0$ and $\alpha>\beta>-1$. So $b_n$ is positive.
{}From this three-term recurrence we obtain the difference equation
for the generating function
$\displaystyle G^{(\alpha,\beta)}_{\rm monic}(t,x,\delta)
=\sum_{n=0}^{\infty}t^nP^{(\alpha,\beta)\,{\rm monic}}_n(x,\delta)$,
\begin{eqnarray}
  \hspace*{-10mm}&&\Bigl((d_{2n-1}d_{2n}^{\ 2}d_{2n+1}d_{2n+2})
  \Bigm|_{n\to t\frac{\partial}{\partial t}}
  (\frac{\partial}{\partial t}+x)
  -(d_{2n-1}d_{2n}^{\ 2}d_{2n+1}d_{2n+2}a_n)
  \Bigm|_{n\to t\frac{\partial}{\partial t}}\n
  \hspace*{-10mm}&&\qquad\qquad\qquad\qquad
  +(d_{2n-1}d_{2n}^{\ 2}d_{2n+1}d_{2n+2}b_n)
  \Bigm|_{n\to t\frac{\partial}{\partial t}}t\,\Bigr)
  G^{(\alpha,\beta)}_{\rm monic}(t,x,\delta)=0\,.
\end{eqnarray}

It is interesting to note that the three-term recurrence
(\ref{rec_rsBCA})--(\ref{rec_rsBC}) simplifies drastically in a `strong'
coupling limit,
\begin{eqnarray}
  \hspace*{-15mm}&&
  g_M\to+\infty \Longleftrightarrow \delta\to1,\qquad g_L,\
  g_S:\ \mbox{fixed},
  \label{stlim1}\\
  \hspace*{-15mm}&&
  a_0={\beta-\alpha\over{\alpha+\beta+2}},\quad
  a_1={(\beta-\alpha)(2\beta+1)\over{(\alpha+\beta+2)(\alpha-\beta+1)}},
  \quad a_n=0,\quad (n\ge2),\\
  \hspace*{-15mm}&&
  b_1=\frac{4(\beta+1)(\alpha+1+(\beta+1)(\alpha-\beta)^2)}
  {(\alpha-\beta+1)(\alpha+\beta+2)^2(2\beta+3)},\ 
  b_2={\alpha+\beta+2\over{2(\alpha-\beta+1)(2\beta+3)}},\ 
  b_n={1\over4},\ (n\ge3).\ \ 
\end{eqnarray}

%%%%%%%%%%%%%
%  C_r      %
%%%%%%%%%%%%%
\paragraph{$C_r$ :}
For the solution $\{\bar{q}_j\}$ of the $C_r$ equation (\ref{rsCeq}),
let us define
\begin{equation}
  \delta=\tanh^2\!g_S,\quad \alpha=\frac{\tanh 2g_L}{2\tanh g_S}-1,
  \label{para_rsC}
\end{equation}
and introduce a degree $r$ polynomial in $x$ having zeros at
$\{\cos 2\bar{q}_j\}$ :
\begin{equation}
  C_r^{(\alpha+\frac12)}(x,\delta)\eqdef
  2^r\schoose{\alpha-\frac12+r}{r}
  \prod_{j=1}^r(x-\cos 2\bar{q}_j)\,.
\label{gegendef}
\end{equation}
It is a deformation of the Gegenbauer polynomial (\ref{Gegendef}) such that
\begin{equation}
  \lim_{\delta\to0}C_r^{(\alpha+\frac12)}(x,\delta)
  =C_r^{(\alpha+\frac12)}(x),
\end{equation}
and obviously it is a special case of $P_r^{(\alpha,\beta)}(x,\delta)$
with definite parity,
\begin{equation}
  C_r^{(\alpha+\frac12)}(x,\delta)
  =\schoose{2\alpha+r}{r}\schoose{\alpha+r}{r}^{-1}
  P_r^{(\alpha,\alpha)}(x,\delta), \quad
C_r^{(\alpha+\frac12)}(-x,\delta)=(-1)^rC_r^{(\alpha+\frac12)}(x,\delta).
\end{equation}
The functional equation  of
$C^{(\alpha+\frac12)}_n(x,\delta)$ is easily obtained from that of
$P^{(\alpha,\beta)}_n(x,\delta)$ and will not be presented here.
The three-term recurrence for $C^{(\alpha+\frac12)}_n(x,\delta)$ reads
\begin{equation}
  \frac{n+1}{2\alpha+2n+1}\,C_{n+1}^{(\alpha+\frac12)}(x,\delta)
  +x\,C_n^{(\alpha+\frac12)}(x,\delta)
  +\frac{2\alpha+2n-1}{n}\, b_n\,C_{n-1}^{(\alpha+\frac12)}(x,\delta)=0\,.
  \label{rec_rsC2}
\end{equation}

%%%%%%%%%%%%%
%  B'_r     %
%%%%%%%%%%%%%
\paragraph{$B'_r$ :}
For the solution $\{\bar{q}_j\}$ of the $B'_r$ equation (\ref{rsBpeq}),
let us define
\begin{equation}
  \delta=\tanh^2\!g_L,\quad \alpha=\frac{\tanh g_S}{\tanh g_L}-1,
  \label{para_rsBp}
\end{equation}
and introduce a degree $r$ polynomial in $x$ having zeros at
$\{\cos 2\bar{q}_j\}$ :
\begin{equation}
  \tilde{P}_r^{(\alpha)}(x,\delta)
  \eqdef 2^{-r}\schoose{\alpha-1+2r}{r}
  \prod_{j=1}^r(x-\cos 2\bar{q}_j)\,.
\end{equation}
Obviously it is a special case of $P_r^{(\alpha,\beta)}(x,\delta)$,
\begin{equation}
  \tilde{P}_r^{(\alpha)}(x,\delta)
  =P_r^{(\alpha,-1)}(x,\delta).
\end{equation}
The functional equation and three-term recurrence of
$\tilde{P}^{(\alpha)}_n(x,\delta)$ are obtained from those of
$P^{(\alpha,\beta)}_n(x,\delta)$.

%%%%%%%%%%%%%%%%%%%%%%%%%%%%
%  5.3                     %
%%%%%%%%%%%%%%%%%%%%%%%%%%%%
\subsection{$C'_r$}
%%%%%%%%%%%%%
%  C'_r     %
%%%%%%%%%%%%%
\paragraph{$C'_r$ :}
For the solution $\{\bar{q}_j\}$ of the $C'_r$ equation (\ref{rsCpeq}),
let us define
\begin{equation}
  \delta=\tanh^2\!g_S,\quad \alpha=\frac{\tanh g_L}{\tanh g_S}-1,
  \label{para_rsCp}
\end{equation}
and introduce a degree $r$ polynomial in $x$ having zeros at
$\{\cos 2\bar{q}_j\}$:
\begin{equation}
  \tilde{C}_r^{(\alpha+\frac12)}(x,\delta)\eqdef
  2^r\schoose{\alpha-\frac12+r}{r}
  \prod_{j=1}^r(x-\cos 2\bar{q}_j)\,.
\label{gegentdef}
\end{equation}
It is a deformation of the Gegenbauer polynomial (\ref{Gegendef})
with definite parity
\begin{equation}
  \lim_{\delta\to0}\tilde{C}_r^{(\alpha+\frac12)}(x,\delta)
  =C_r^{(\alpha+\frac12)}(x),\quad
\tilde{C}_r^{(\alpha+\frac12)}(-x,\delta)
  =(-1)^r\tilde{C}_r^{(\alpha+\frac12)}(x,\delta).
\end{equation}

The functional equation for $\tilde{C}^{(\alpha+\frac12)}_n(x,\delta)$ 
($n\geq 1$) reads
\begin{eqnarray}
  &&\phantom{-}\Bigl(\dr\,x-i\sqrt{1-x^2}\Bigr)
  \Bigl((\alpha+1)\dr\,x+i\sqrt{1-x^2}\Bigr)^{\!2}
  \tilde{C}_n^{(\alpha+\frac12)}
  \Bigl(\sfrac{1+\delta}{1-\delta}\,x
  +i\sfrac{2\dr}{1-\delta}\sqrt{1-x^2},\delta\Bigr)\n
  &&-
  \Bigl(\dr\,x+i\sqrt{1-x^2}\Bigr)
  \Bigl((\alpha+1)\dr\,x-i\sqrt{1-x^2}\Bigr)^{\!2}
  \tilde{C}_n^{(\alpha+\frac12)}
  \Bigl(\sfrac{1+\delta}{1-\delta}\,x
  -i\sfrac{2\dr}{1-\delta}\sqrt{1-x^2},\delta\Bigr)\n
  &&=
  2i\sqrt{1-x^2}\,
  (A_nx^2+B_nx+C_n)\tilde{C}_n^{(\alpha+\frac12)}(x,\delta).
  \label{feq_rsCp}
\end{eqnarray}
Here $A_n=A^{(\alpha+\frac12)}_n(\delta)$,
$B_n=B^{(\alpha+\frac12)}_n(\delta)$ and
$C_n=C^{(\alpha+\frac12)}_n(\delta)$ are real constants:
\begin{eqnarray}
  A_n&\!\!=\!\!&
  -(1-\delta\,)^{-(n-1)}\times
   \frac12\Bigl((1+(\alpha+1)\dr\,)^2(1+\dr\,)^{2n-1}\n
  &&\qquad\qquad\qquad\qquad\quad
     +(1-(\alpha+1)\dr\,)^2(1-\dr\,)^{2n-1}\Bigr),\\
  B_n&\!\!=\!\!&0\,,\\
  C_n&\!\!=\!\!&
  (1-\delta\,)^{-n}\times
   \frac12\Bigl((1+(\alpha+1)\dr\,)^2(1+\dr\,)^{2n-1}\n
  &&\qquad\qquad\qquad
     +(1-(\alpha+1)\dr\,)^2(1-\dr\,)^{2n-1}
     -2\alpha^2\delta\,(1-\delta\,)^{n-1}\Bigr).\quad
  \label{feq_rsCp_coef}
\end{eqnarray}
This functional equation contains all the information of the equilibrium.

The deformed Gegenbauer polynomial 
$\tilde{C}^{(\alpha+\frac12)}_n(x,\delta)$ satisfies the three-term 
recurrence (\ref{3termrec}) (see Appendix) with
\begin{eqnarray}
  a_n&\!\!=\!\!&0\,,
  \label{recCpsta}\\
  b_n&\!\!=\!\!&
  \frac{1}{2\dr}\Bigl((1+\dr\,)^n-(1-\dr\,)^n\Bigr) \n
  &&\times
  \frac{1}{2^2}\Bigl((1+(\alpha+1)\dr\,)(1+\dr\,)^{n-1}
     +(1-(\alpha+1)\dr\,)(1-\dr\,)^{n-1}\Bigr)^{\!2}\n
  &&\times
  \frac{d_{n}}{d_{2n-1}\,d_{2n+1}},
  \quad c_n=2^n\schoose{\alpha-\frac12+n}{n},
  \label{rec_rsCp}\\
  d_m&\!\!=\!\!&\frac{1}{2\dr}\Bigl(
  (1+(\alpha+1)\dr\,)^2(1+\dr\,)^{m-2}
  -(1-(\alpha+1)\dr\,)^2(1-\dr\,)^{m-2}\Bigr).
  \label{recCpstd}
\end{eqnarray}
Namely $\tilde{C}^{(\alpha+\frac12)}_n(x,\delta)$ satisfies
\begin{equation}
  \frac{n+1}{2\alpha+2n+1}\tilde{C}_{n+1}^{(\alpha+\frac12)}(x,\delta)
  +x\,\tilde{C}_n^{(\alpha+\frac12)}(x,\delta)
  +\frac{2\alpha+2n-1}{n}\,b_n\,
  \tilde{C}_{n-1}^{(\alpha+\frac12)}(x,\delta)=0.
  \label{rec_rsCp2}
\end{equation}
In this case (\ref{para_rsCp}), the parameters are
$\delta>0$ and $\alpha>-1$. So $b_n$ is positive.
{}From this three-term recurrence we obtain the difference equation
for the generating function
$\displaystyle G^{(\alpha+\frac12)}(t,x,\delta)
=\sum_{n=0}^{\infty}t^n\tilde{C}^{(\alpha+\frac12)}_n(x,\delta)$
in a similar way to that of  $P^{(\alpha,\beta)}_n(x,\delta)$.

\bigskip

In a `strong' coupling limit
\begin{equation}
  g_S\to+\infty\Longleftrightarrow \delta\to1,\qquad g_L : \mbox{fixed},
\label{stlim2}
\end{equation}
the three-term recurrence (\ref{rec_rsCp})--(\ref{recCpstd}) simplifies
drastically:
\begin{eqnarray}
  b_1={1\over{(\alpha+2)^2}},\quad
  b_2={\alpha+1\over{(\alpha+2)^2}},\quad
  b_n={1\over4},\quad (n\ge3).
\end{eqnarray}

%%%%%%%%%%%%%%%%%%%%%%%%%%%%
%  5.4                     %
%%%%%%%%%%%%%%%%%%%%%%%%%%%%
\subsection{$B_r$}
%%%%%%%%%%%%%
%  B_r      %
%%%%%%%%%%%%%
\paragraph{$B_r$ :}
For the solution $\{\bar{q}_j\}$ of the $B_r$ equation (\ref{rsBeq}),
let us define
\begin{equation}
  \delta=\tanh^2\!g_L,\quad \alpha=\frac{2\tanh\frac{g_S}{2}}{\tanh g_L}-1,
  \label{para_rsB}
\end{equation}
and introduce a degree $r$ polynomial in $x$ having zeros at
$\{\cos 2\bar{q}_j\}$:
\begin{equation}
  \hat{P}_r^{(\alpha)}(x,\delta)
  \eqdef 2^{-r}\schoose{\alpha-1+2r}{r}
  \prod_{j=1}^r(x-\cos 2\bar{q}_j)\,.
\label{phatdef}
\end{equation}
It is a deformation of the Jacobi polynomial (\ref{Pdef}) such that
\begin{equation}
  \lim_{\delta\to 0}\hat{P}_r^{(\alpha)}(x,\delta)=P_r^{(\alpha,-1)}(x).
\end{equation}

The functional equation for $\hat{P}^{(\alpha)}_n(x,\delta)$ 
($n\geq 1$) reads
\begin{eqnarray}
  &&\phantom{+}\Bigl(\dr\,x-i\sqrt{1-x^2}\Bigr)
  \Bigl(\sfrac{\alpha+1}{2}\dr\,(1+x)+i\sqrt{1-x^2}\Bigr)^{\!2}
  \hat{P}_n^{(\alpha)}
  \Bigl(\sfrac{1+\delta}{1-\delta}\,x
  +i\sfrac{2\dr}{1-\delta}\sqrt{1-x^2},\delta\Bigr)\n
  &&-
  \Bigl(\dr\,x+i\sqrt{1-x^2}\Bigr)
  \Bigl(\sfrac{\alpha+1}{2}\dr\,(1+x)-i\sqrt{1-x^2}\Bigr)^{\!2}
  \hat{P}_n^{(\alpha)}
  \Bigl(\sfrac{1+\delta}{1-\delta}\,x
  -i\sfrac{2\dr}{1-\delta}\sqrt{1-x^2},\delta\Bigr)\n
  &&=
  2i\sqrt{1-x^2}\,(A_nx^2+B_nx+C_n)\hat{P}_n^{(\alpha)}(x,\delta).
  \label{feq_rsB}
\end{eqnarray}
Here $A_n=A^{(\alpha)}_n(\delta)$,
$B_n=B^{(\alpha)}_n(\delta)$ and
$C_n=C^{(\alpha)}_n(\delta)$ are real constants:
\begin{eqnarray}
  A_n&\!\!=\!\!&
  -(1-\delta\,)^{-(n-1)}\times
   \frac12\Bigl((1+\sfrac12(\alpha+1)\dr\,)^2(1+\dr\,)^{2n-1}\n
  &&\qquad\qquad\qquad\qquad
     +(1-\sfrac12(\alpha+1)\dr\,)^2(1-\dr\,)^{2n-1}\Bigr),\\
  B_n&\!\!=\!\!&
  \sfrac12(1-\alpha^2)\delta\,,\\
  C_n&\!\!=\!\!&
  (1-\delta\,)^{-n}\times
   \frac12\Bigl((1+\sfrac12(\alpha+1)\dr\,)^2(1+\dr\,)^{2n-1}\n
  &&\qquad\qquad\qquad
     +(1-\sfrac12(\alpha+1)\dr\,)^2(1-\dr\,)^{2n-1}\n
  &&\qquad\qquad\qquad
     -(1+\alpha^2-\sfrac12(\alpha+1)^2\delta)\delta\,
     (1-\delta\,)^{n-1}\Bigr).\quad
  \label{feq_rsB_coef}
\end{eqnarray}
This functional equation contains all the information of the equilibrium.

The deformed Jacobi polynomial $\hat{P}^{(\alpha)}_n(x,\delta)$
satisfies the three-term recurrence (\ref{3termrec}) (see Appendix) with
\begin{eqnarray}
  a_n&\!\!=\!\!&
  (1-\alpha^2)(1-\delta)^{n-1}
  \frac{d_n^{\,\prime}\,d_{n+1}^{\,\prime}}{d_{2n}\,d_{2n+2}}\,,
  \label{recBsta}\\
  b_n&\!\!=\!\!&
  4\,\frac{1}{2\dr}\Bigl((1+\dr\,)^{n-1}-(1-\dr\,)^{n-1}\Bigr)
  \times\frac{1}{2\dr}\Bigl((1+\dr\,)^n-(1-\dr\,)^n\Bigr) \n
  &&\times
  \frac{d_n^{\,\prime\:4}\,d_n\,d_{n+1}}{d_{2n-1}\,d_{2n}^{\ 2}\,d_{2n+1}},
  \quad c_n=2^{-n}\schoose{\alpha-1+2n}{n},
  \label{rec_rsB}\\
  d_m&\!\!=\!\!&\frac{1}{2\dr}\Bigl(
  (1+\sfrac12(\alpha+1)\dr\,)^2(1+\dr\,)^{m-2}
  -(1-\sfrac12(\alpha+1)\dr\,)^2(1-\dr\,)^{m-2}\Bigr),\quad\\
  d_m^{\,\prime}&\!\!=\!\!&\frac{1}{2}\Bigl(
  (1+\sfrac12(\alpha+1)\dr\,)(1+\dr\,)^{m-1}
  +(1-\sfrac12(\alpha+1)\dr\,)(1-\dr\,)^{m-1}\Bigr).
  \label{recBstd}
\end{eqnarray}
Namely $\hat{P}^{(\alpha)}_n(x,\delta)$ satisfies
\begin{eqnarray}
  &&\frac{2(n+1)(\alpha+n)}{(\alpha+2n)(\alpha+2n+1)}
  \hat{P}_{n+1}^{(\alpha)}(x,\delta)
  +(x-a_n)\hat{P}_n^{(\alpha)}(x,\delta)\n
  &&\qquad\qquad\qquad\qquad
  +\frac{(\alpha+2n-2)(\alpha+2n-1)}{2n(\alpha+n-1)}\,
  b_n\,\hat{P}_{n-1}^{(\alpha)}(x,\delta)=0\,,
  \label{rec_rsB2}
\end{eqnarray}
In this case (\ref{para_rsB}), the parameters are
$\delta>0$ and $\alpha>-1$. So $b_n$ is positive.
{}From this three-term recurrence we obtain the difference equation
for the generating function
$\displaystyle G^{(\alpha)}(t,x,\delta)
=\sum_{n=0}^{\infty}t^n\hat{P}^{(\alpha)}_n(x,\delta)$
in a similar way to that of $P^{(\alpha,\beta)}_n(x,\delta)$.

In a `strong' coupling limit
\begin{equation}
  g_L\to+\infty\Longleftrightarrow \delta\to1,\qquad g_S : \mbox{fixed},
\label{stlim3}
\end{equation}
the three-term recurrence (\ref{recBsta})--(\ref{recBstd}) simplifies
drastically:
\begin{eqnarray}
  &&a_0=-1,\quad a_1={1-\alpha\over{\alpha+3}},\quad
  a_n=0,\quad (n\ge2),\\
  &&b_1=0,\quad
  b_2={2(\alpha+1)\over{(\alpha+3)^2}},\quad
  b_n={1\over4},\quad (n\ge3).
\end{eqnarray}

%%%%%%%%%%%%%
%  D_r      %
%%%%%%%%%%%%%
\paragraph{$D_r$ :}
As in the Sutherland systems, the $D_r$ is a special case $g_S=0$ of
the $B_r$ theory described by $\hat{P}_r^{(-1)}(x,\delta)$,
which has a zero at $x=\pm 1$ for $r\geq 2$.

%%%%%%%%%%%%%
%  identity %
%%%%%%%%%%%%%
\paragraph{Deformation of the identities :}
Before closing this section let us briefly discuss the
deformation of the identities between the Chebyshev and Jacobi 
polynomials (\ref{chebjaciden1})--(\ref{chebjaciden2})
and those between the Jacobi polynomials (\ref{gegiden}).
As remarked in section \ref{nodeform}, the Chebyshev polynomials
describing the equilibrium of the $A_{r-1}$ systems are not 
{\em deformed\/} in our scheme. Therefore no deformation of the 
identities (\ref{chebjaciden1})--(\ref{chebjaciden2}) exists.
The folding $D_{r+1}\to B_r$ (\ref{DtoB}) leads to the identity 
between the deformed Jacobi polynomials $\hat{P}_{r}^{(\alpha)}(x,\delta)$
(\ref{phatdef}) associated with the $B_r$ systems
\begin{equation}
  2(r+1)\hat{P}_{r+1}^{(-1)}(x,\delta)
  =r(x-1) \hat{P}_r^{(1)}(x,\delta)\,,
\label{Jacidendef}
\end{equation}
which is a deformation of the identity (\ref{gegiden}).
As remarked at the end of section \ref{cssub}, we have not been able to 
deform the identities between the Gegenbauer and Jacobi polynomials
(\ref{genjaciden1})--(\ref{genjaciden2}) as they do not seem to have a
root theoretic explanation.

Among the deformed Jacobi polynomials
$P^{(\alpha,\beta)}_n(x,\delta)$ for various $\alpha$ and $\beta$,
two cases (i) $\alpha=\beta=-1/2$, (ii) $\alpha=-\beta=1/2$ are
not {\em deformed\/}:
\begin{equation}
  P^{(-1/2,-1/2)}_n(x,\delta)=P^{(-1/2,-1/2)}_n(x),\quad
  P^{(1/2,-1/2)}_n(x,\delta)=P^{(1/2,-1/2)}_n(x).
\end{equation}
The first is the Chebyshev polynomial of the first kind
$T_n(x)\propto \cos n\varphi$, $x=\cos\varphi$, as remarked in
section \ref{nodeform}.
The second case is $P^{(1/2,-1/2)}_n(x)\propto
\sin((2n+1)\varphi/2)/\sin(\varphi/2)$, $x=\cos\varphi$.
In both cases, the zeros of $P^{(\alpha,\beta)}_n(x)$ are
{\em equally spaced\/}.
There is a third case \cite{cs,szego} of equally spaced zeros of
the Jacobi polynomial $P^{(\alpha,\beta)}_n(x)$, for $\alpha=\beta=1/2$,
corresponding to the Chebyshev polynomial of the second kind 
$U_n(x)\propto \sin n\varphi/\sin\varphi$, $x=\cos\varphi$, which is, 
interestingly, {\em deformed\/}. We have no explanation to offer.

%%%%%%%%%%%%%%%%%%%%%%%%%%%%%%%%%%%%%%%%%%%%%%%%%%%%%%%%%%%%%%%
%                                                             %
%  6. Summary and Comments                                    %
%                                                             %
%%%%%%%%%%%%%%%%%%%%%%%%%%%%%%%%%%%%%%%%%%%%%%%%%%%%%%%%%%%%%%%
\section{Summary and Comments}
\setcounter{equation}{0}

We have derived certain deformation of the classical orthogonal
polynomials (the Hermite, Laguerre, Gegenbauer and Jacobi)
describing the equilibrium of a class of multiparticle dynamics,
the Ruijsenaars-Schneider systems.
The R-S systems are `good' deformation of the Calogero and Sutherland
systems whose equilibrium points are described by the zeros
of the above classical orthogonal polynomials.
As remarked in the text these deformed polynomials do not belong to 
the $q$-deformed orthogonal polynomials \cite{And-Ask-Roy}
or their analogs \cite{jingyang}.

The quality and quantity of the knowledge of these new polynomials are
rather varied.
The one parameter deformation of the Hermite polynomials section 
\ref{sec:H.1} is best understood.
Its three-term recurrence (\ref{rec_rclA2})  tells that it is the simplest
possible deformation which reduces to the original Hermite polynomial
without rescaling {\em etc\/} in the zero deformation limit
($\delta\to0$). As shown in some detail in section \ref{sec:H.1}, the
generating function (\ref{G_ans}) and the weight function
(\ref{wint}) are known explicitly. Some identities connecting the
Hermite and Laguerre polynomials (\ref{LHiden})--(\ref{LHiden2})
are nicely deformed (\ref{LHidendef1})--(\ref{LHidendef2}),
(\ref{LHiden2def1})--(\ref{LHiden2def2}). It is interesting to
note that some non-trivial identities of the Hermite polynomials
are preserved after deformation. For example, the `addition
theorem' reads
\begin{equation}
  \sum_{n_1,\cdots,n_m=0\atop n_1+\cdots+n_m=n}^{\infty}
  \frac{\alpha_1^{n_1}\cdots\alpha_m^{n_m}}{n_1!\cdots n_m!}
  H_{n_1}(x_1)\cdots H_{n_m}(x_m)
  =\frac{|\vec{\alpha}|^n}{n!}
  H_n\Bigl(\frac{\vec{\alpha}\cdot\vec{x}}{|\vec{\alpha}|}\Bigr),
\end{equation}
in which the following notation is used:
$\vec{\alpha}={}^t(\alpha_1,\cdots,\alpha_m)$,
$\vec{x}={}^t(x_1,\cdots,x_m)$,
$|\vec{\alpha}|=\sqrt{\alpha_1^2+\cdots+\alpha_m^2}$,\
$\vec{\alpha}\cdot\vec{x}=\alpha_1x_1+\cdots+\alpha_mx_m$.
The deformed version is
\begin{equation}
  \sum_{n_1,\cdots,n_m=0\atop n_1+\cdots+n_m=n}^{\infty}
  \frac{\alpha_1^{n_1}\cdots\alpha_m^{n_m}}{n_1!\cdots n_m!}
  H_{n_1}\Bigl(x_1,\frac{\delta}{\alpha_1^2}\Bigr)\cdots
  H_{n_m}\Bigl(x_m,\frac{\delta}{\alpha_m^2}\Bigr)
  =\frac{|\vec{\alpha}|^n}{n!}
  H_n\Bigl(\frac{\vec{\alpha}\cdot\vec{x}}{|\vec{\alpha}|},
  \frac{\delta}{|\vec{\alpha}|^2}\Bigr).
\end{equation}
Both can be derived from the generating functions.

After a long and laborious search through existing literature, we find out
that all the deformed orthogonal polynomials introduced in the main
text can be related to particular members of the Askey-scheme of
hypergeometric orthogonal polynomials \cite{koeswart}.

The deformed Hermite polynomial $H_n(x,\delta)$ is related to the
Meixner-Pollaczek polynomial $P_n^{(\lambda)}(x;\phi)$
(\S1.7 of \cite{koeswart}):
\begin{equation}
  H_n(x,\delta)=n!\dr^{\,n}P_n^{(\frac{1}{\delta})}
  \Bigl(\frac{x}{\dr}\,;\frac{\pi}{2}\Bigr).
\end{equation}
This identification allows a simple expression of its general term
in terms of a
(truncated) hypergeometric series ${}_2F_1$:
\begin{eqnarray}
  H_n(x,\delta)&\!\!=\!\!&i^n\dr^{\,n}\Bigl(\frac{\delta}{2}\Bigr)_n\,
  {}_2F_1\Bigl({-n,\frac{1}{\delta}+i\frac{x}{\dr}\atop\frac{2}{\delta}}
  \Bigm|2\Bigr)\n
  &\!\!=\!\!&
  2^ni^n\sum_{k=0}^n(-1)^k{n\choose k}
  \prod_{j=0}^{k-1}\Bigl(ix+\frac{1}{\dr}+\dr\,j\Bigr)\times
  \prod_{j=k}^{n-1}\Bigl(\frac{1}{\dr}+\frac{\dr}{2}j\Bigr),
\end{eqnarray}
which is deformation of (\ref{hermdef}).
The two-parameter deformation of the Hermite polynomial
$H_n(x,\delta,\varepsilon)$ (\ref{twoherm}) is a special case of the
continuous Hahn polynomial (\S1.4 of \cite{koeswart}):
\begin{equation}
  H_n(x,\delta,\varepsilon)=\frac{2^nn!\dr^{\,n}}
  {(n-1+\frac{2}{\delta}+\frac{2}{\delta\varepsilon})_n}
  \,p_n\Bigl(\frac{x}{\dr}\,;\frac{1}{\delta}\,,\frac{1}{\delta\varepsilon}\,,
  \frac{1}{\delta}\,,\frac{1}{\delta\varepsilon}\Bigr).
\end{equation}
The (two-parameter) deformed Laguerre polynomial
$L_n^{(\alpha)}(x,\gamma,\delta)$ (\ref{twodefLag}) is
the continuous dual Hahn polynomial (\S1.3 of \cite{koeswart}) with rescaling:
\begin{equation}
  L_n^{(\alpha)}(y^2,\gamma,\delta)=\frac{\delta^n}{n!}
  S_n\Bigl(\frac{y^2}{\delta}\,;
  \frac{1}{\delta}\,,\alpha_1+1,\alpha_2+1\Bigr),
\end{equation}
in which $\alpha_1$ and $\alpha_2$ are the two roots of
$x^2-(\alpha-1)x+\gamma-\alpha=0$.
The (three-parameter) deformed Laguerre polynomial
$L_n^{(\alpha)}(x,\gamma,\delta,\varepsilon)$ (\ref{threedefLag}) is the
Wilson polynomial (\S1.1 of \cite{koeswart}) with rescaling:
\begin{equation}
  L_n^{(\alpha)}(y^2,\gamma,\delta,\varepsilon)
  =\frac{\delta^{\,n}}
  {n!\,(n+\alpha+\frac{1}{\delta}+\frac{1}{\delta\varepsilon})_n}
  W_n\Bigl(\frac{y^2}{\delta}\,;\frac{1}{\delta}\,,
  \frac{1}{\delta\varepsilon}\,,\alpha_1+1,\alpha_2+1\Bigr).
\end{equation}
The deformed Jacobi polynomial $P_n^{(\alpha,\beta)}(x,\delta)$
(\ref{defjacobidef})
is a special case of the Askey-Wilson polynomial (\S3.1 of \cite{koeswart}):
\begin{eqnarray}
  &&P_n^{(\alpha,\beta)}(x,\delta)
  =2^{-2n}\schoose{\alpha+\beta+2n}{n}(ab^2q^{n-1};q)_n^{-1}\,
  p_n(x\,;a,b,-b,-1\,|q),
  \label{aw1}\\[6pt]
  &&q=\frac{1-\dr}{1+\dr}=e^{-2g_M},\ \
  a=\frac{1-(\alpha-\beta)\dr}{1+(\alpha-\beta)\dr}=e^{-2g_S},\ \
  b^2=\frac{1-2(\beta+1)\dr}{1+2(\beta+1)\dr}=e^{-4g_L}.\qquad
\end{eqnarray}
The deformed Gegenbauer polynomial
$\tilde{C}_n^{(\alpha+\frac12)}(x,\delta)$ (\ref{gegentdef})
is again a special case of the Askey-Wilson polynomial (\S3.1 of
\cite{koeswart}):
\begin{eqnarray}
  &&\tilde{C}_n^{(\alpha+\frac12)}(x,\delta)
  =\schoose{\alpha-\frac12+n}{n}(a^4q^{n-1};q)_n^{-1}\,
  p_n(x\,;a,a,-a,-a\,|q),
  \label{aw2}\\
  &&q=\frac{1-\dr}{1+\dr}=e^{-2g_S},\qquad
  a^2=\frac{1-(\alpha+1)\dr}{1+(\alpha+1)\dr}=e^{-2g_L}.
\end{eqnarray}
Another deformation of the Jacobi polynomial
$\hat{P}_n^{(\alpha)}(x,\delta)$ (\ref{phatdef}) is also a special case 
of the Askey-Wilson polynomial (\S3.1 of \cite{koeswart}):
\begin{eqnarray}
  &&\hat{P}_n^{(\alpha)}(x,\delta)
  =2^{-2n}\schoose{\alpha-1+2n}{n}(a^2q^{n-1};q)_n^{-1}\,
  p_n(x\,;a,a,-1,-1|q),
  \label{aw3}\\
  &&q=\frac{1-\dr}{1+\dr}=e^{-2g_L},\qquad
  a=\frac{1-\frac12(\alpha+1)\dr}{1+\frac12(\alpha+1)\dr}=e^{-g_S}.
\end{eqnarray}
In all these formulas the Pochhammer symbol
$(a)_k=\prod_{j=0}^{k-1}(a+j)$ and its $q$-extension 
$(a;q)_k=\prod_{j=0}^{k-1}(1-aq^j)$ are used.

Various `strong' coupling limits (\ref{stlim1}), (\ref{stlim2}) and
(\ref{stlim3}) of the deformed Jacobi type polynomials
$P_n^{(\alpha,\beta)}(x,\delta)$,
$\tilde{C}_n^{(\alpha+\frac12)}(x,\delta)$ and
$\hat{P}_n^{(\alpha)}(x,\delta)$ simply correspond to the `{\em crystal\/}'
limit $q\to0+$ of the Askey-Wilson polynomials (\ref{aw1}), (\ref{aw2}) and
(\ref{aw3}).

For all these polynomials discussed in the present paper, the general 
term can be expressed in terms of various hypergeometric functions
${}_2F_1$, ${}_3F_2$, ${}_4F_3$ and ${}_4\phi_3$. A Rodrigue type
formula, the generating function and the weight function {\em etc\/} 
can be written down by using the general formulas of the Askey-scheme 
of hypergeometric orthogonal polynomials \cite{koeswart}.

As remarked repeatedly in the text, the equations determining the 
equilibrium positions, (\ref{rclAeq})--(\ref{rclDeq}), 
(\ref{rcqAeq})--(\ref{rcqDeq}) and (\ref{rsAeq})--(\ref{rsDeq}) look 
similar to the Bethe ansatz equation.
For the simplest spin $1/2$ XXX  chain with $N$ sites,
the Bethe ansatz equation reads
\begin{equation}
  \prod_{k=1\atop k\neq j}^l
  {\bar{q}_j-\bar{q}_k+2i\over{\bar{q}_j-\bar{q}_k-2i}}
  =\left({\bar{q}_j+i\over{\bar{q}_j-i}}\right)^{\!N},\quad
  (j=1,\ldots,l),
  \label{Beq}
\end{equation}
in which $l\le N$ is the number of up (down) spins.
This looks similar to the rational $A$ type equations (\ref{rclAeq})
and (\ref{rcqAeq}) with a special choice of the potential $w(x)$ function,
the $N$-th power rather than linear or quadratic.
The corresponding functional relation, Baxter's t-Q relation,
looks very much like the functional equations 
(\ref{feq_rclA}) and (\ref{feq_rcqA}).
It would be interesting to pursue the analogy further.

%%%%%%%%%%%%%%%%%%%%%%%%%%%%%%%%%%%%%%%%%%%%%%%%%%%%%%%%%%%%%%%
%                                                             %
%  Acknowledgments                                            %
%                                                             %
%%%%%%%%%%%%%%%%%%%%%%%%%%%%%%%%%%%%%%%%%%%%%%%%%%%%%%%%%%%%%%%
\section*{Acknowledgements}

We thank M.\, Rossi and K.\, Hikami for useful discussion.
S. O. and R. S. are supported in part by Grant-in-Aid for Scientific
Research from the Ministry of Education, Culture, Sports, Science and
Technology, No.13135205 and No. 14540259, respectively and S. O.
is also supported by `Gakubucho Sairyou Keihi' of Faculty of Science,
Shinshu University.

%%%%%%%%%%%%%%%%%%%%%%%%%%%%%%%%%%%%%%%%%%%%%%%%%%%%%%%%%%%%%%%
%                                                             %
%  Appendix: Relation between the Functional Equation and     %
%            the Three Term Recurrence                        %
%                                                             %
%%%%%%%%%%%%%%%%%%%%%%%%%%%%%%%%%%%%%%%%%%%%%%%%%%%%%%%%%%%%%%%
\renewcommand{\thesection}{\Alph{section}}
\setcounter{section}{1}
\renewcommand{\theequation}{A.\arabic{equation}}
\setcounter{equation}{0}
\section*{Appendix: Relation between the Functional Equation and the
Three Term Recurrence}

In this appendix we show the relation between the functional equation
and the three-term recurrence, without proof.
Since the normalisation of polynomials is irrelevant, we consider
monic polynomials $f_n(x)$ with real coefficients and without
superscript `monic'.

The three-term recurrence of $f_n(x)$ (\ref{3termrec}) is
\begin{equation}
  f_{n+1}(x)=(x-a_n)f_n(x)-b_nf_{n-1}(x)\,,\quad(n\geq 0)\,,
  \label{3termrecf}
\end{equation}
with $f_{-1}(x)=0$ and $f_0(x)=1$.
The explicit forms of $a_n$ and $b_n$ can be
read from (\ref{rec_rclA}), (\ref{rec_rcqA}),
(\ref{rec_rclBCt0})--(\ref{rec_rclBCt}),
(\ref{rec_rcqBCt0})--(\ref{rec_rcqBCt}),
(\ref{rec_rsBCA})--(\ref{rec_rsBC}),
(\ref{recCpsta})--(\ref{rec_rsCp}),
(\ref{recBsta})--(\ref{rec_rsB}).

The functional equations for the deformed Hermite, Laguerre and Jacobi
polynomials have the following forms ($n\geq -1$):
\begin{eqnarray}
  \mbox{Hermite}&:&
  h(x)f_n(x+i\dr\,)+\epsilon\,h(x)^* f_n(x-i\dr\,)
  =2\,i^{\frac{1-\epsilon}{2}}g_n(x)f_n(x)\,,
  \label{feq_H}\\
  \mbox{Laguerre}&:&
  h(y)f_n\Bigl((y+i\dr\,)^2\Bigr)
  +\epsilon\,h(y)^* f_n\Bigl((y-i\dr\,)^2\Bigr)
  =2\,i^{\frac{1-\epsilon}{2}}g_n(y)f_n(y^2)\,,
  \label{feq_L}\\
  \mbox{Jacobi}&:&
  h(x)f_n\Bigl(\sfrac{1+\delta}{1-\delta}\,x
  +i\,\sfrac{2\dr}{1-\delta}\sqrt{1-x^2}\Bigr)
  +\epsilon\,h(x)^*f_n\Bigl(\sfrac{1+\delta}{1-\delta}\,x
  -i\,\sfrac{2\dr}{1-\delta}\sqrt{1-x^2}\Bigr)\n
  &&\quad=
  2\Bigl(i\sqrt{1-x^2}\,\Bigr)^{\frac{1-\epsilon}{2}}
  g_n(x)f_n(x)\,,
  \label{feq_J}
\end{eqnarray}
where $\epsilon$ is
\begin{eqnarray*}
  \epsilon=-1&:&H_n(x,\delta),
  L^{(\alpha)}_n(x,\gamma,\delta),
  L^{(\alpha)}_n(x,\delta),
  \tilde{L}^{(\alpha)}_n(x,\delta),
  P^{(\alpha,\beta)}_n(x,\delta),
  \tilde{C}^{(\alpha+\frac12)}_n(x,\delta),
  \hat{P}^{(\alpha)}_n(x,\delta),\\
  \epsilon=1&:&H_n(x,\delta,\varepsilon),
  L^{(\alpha)}_n(x,\gamma,\delta,\varepsilon),
  L^{(\alpha)}_n(x,\delta,\varepsilon),
  \tilde{L}^{(\alpha)}_n(x,\delta,\varepsilon),
  C^{(\alpha+\frac12)}_n(x,\delta),
  \tilde{P}^{(\alpha)}_n(x,\delta).
\end{eqnarray*}
The explicit forms of $h(x)$ and $g_n(x)$ can be
read from (\ref{feq_rclA}), (\ref{feq_rcqA}),
(\ref{feq_rclBCt}), (\ref{feq_rcqBCt}),
(\ref{feq_rsBC}), (\ref{feq_rsCp}) and (\ref{feq_rsB})\footnote{
The explicit forms of the function $h(x)$ are derived from the equations
for the equilibrium.
For the deformed Hermite polynomials, $g_n(x)$ are determined by the
consistency of this functional equation.
For the deformed Laguerre and Jacobi polynomials, however, $g_n(x)$ are
not determined uniquely by these functional equations. We have fixed
$g_n(x)$ by using some empirical knowledge of their lower degree members.
}.

As the first step, we show the following property of the functional
equations:
\begin{prop}\label{prop:1}
The solution of the functional equation
{\rm (\ref{feq_H})}--{\rm (\ref{feq_J})}, if exists,
is unique up to an overall normalisation.
\end{prop}
By this proposition, it is sufficient to construct one solution of
each functional equation explicitly.
We will do this by using the three-term recurrence.

As the second step, we show the following relation:
\begin{prop}\label{prop:3termrec2}
Three-term recurrence {\rm (\ref{3termrecf})} implies the
relation {\rm($n\geq 0$)}\footnote{
Although $a_n$ vanishes for the deformed Hermite cases,
we keep $a_n$ in this generic formula.
}
\begin{eqnarray}
  {\rm Hermite}:&&
  i^{\frac{1+\epsilon}{2}}\dr\,h(x)f_n(x+i\dr\,)\n
  &=\!\!&\Bigl((x-a_n)(g_{n+1}(x)-g_n(x))+i\dr\,g_n(x)\Bigr)
  f_n(x)\n
  &&\qquad -b_n\Bigl(g_{n+1}(x)-g_{n-1}(x)\Bigr)f_{n-1}(x)\,,
  \label{3termrec2_H}\\
  {\rm Laguerre}:&&
  2i^{\frac{1+\epsilon}{2}}\dr\,yh(y)f_n\Bigl((y+i\dr\,)^2\Bigr)\n
  &=\!\!&
  \Bigl((y^2-a_n)(g_{n+1}(y)-g_n(y))+\delta g_n(y)+2i\dr\,yg_n(y)\Bigr)
  f_n(y^2)\n
  &&\qquad -b_n(g_{n+1}(y)-g_{n-1}(y))f_{n-1}(y^2)\,,
  \label{3termrec2_L}\\
  {\rm Jacobi}:&&
  \Bigl(i\sqrt{1-x^2}\,\Bigr)^{\frac{1+\epsilon}{2}}
  \sfrac{2\dr}{1-\delta}\,h(x)
  f_n\Bigl(\sfrac{1+\delta}{1-\delta}\,x
  +i\,\sfrac{2\dr}{1-\delta}\sqrt{1-x^2}\Bigr)\n
  &\!\!=\!\!&
  \Bigl((x-a_n)g_{n+1}(x)-(\sfrac{1+\delta}{1-\delta}\,x-a_n)g_n(x)
  +2i\sqrt{1-x^2}\,\sfrac{\dr}{1-\delta}\,g_n(x)\Bigr)f_n(x)\n
  &&\qquad -b_n(g_{n+1}(x)-g_{n-1}(x))f_{n-1}(x)\,.
  \label{3termrec2_J}
\end{eqnarray}
\end{prop}
In the proof of this proposition by induction,
we encounter the equation,
\begin{equation}
  X_n(x)f_n(x)-Y_n(x)b_nf_{n-1}(x)\stackrel{?}{=}0 \quad\mbox{or}\quad
  X_n(y)f_n(y^2)-Y_n(y)b_nf_{n-1}(y^2)\stackrel{?}{=}0.
\end{equation}
Here $X_n(x)$ and $Y_n(x)$ are as follows:\\
Hermite:
\begin{eqnarray}
  X_n(x)&\!\!=\!\!&(x-a_n)^2(g_{n+1}(x)-g_n(x))
  -(x-a_n)(x-a_{n+1})(g_{n+2}(x)-g_{n+1}(x))-\delta g_n(x)\n
  &&
  -b_n(g_n(x)-g_{n-2}(x))+b_{n+1}(g_{n+2}(x)-g_n(x))\,,\\
  Y_n(x)&\!\!=\!\!&(x-a_n)(g_{n+1}(x)-g_{n-1}(x))
  -(x-a_{n-1})(g_{n-1}(x)-g_{n-2}(x))\n
  &&
  -(x-a_{n+1})(g_{n+2}(x)-g_{n+1}(x)),
\end{eqnarray}
Laguerre :
\begin{eqnarray}
  X_n(y)&\!\!=\!\!&(y^2-a_n-\delta)
  \Bigl((y^2-a_n)(g_{n+1}(y)-g_n(y))+\delta g_n(y)\Bigr)\n
  &&
  -(y^2-a_n)\Bigl((y^2-a_{n+1})(g_{n+2}(y)-g_{n+1}(y))
  +\delta g_{n+1}(y)\Bigr)-4\delta\,y^2g_n(y)\n
  &&
  -b_n(g_n(y)-g_{n-2}(y))+b_{n+1}(g_{n+2}(y)-g_n(y))\,,\\
  Y_n(y)&\!\!=\!\!&(y^2-a_n-2\delta)(g_{n+1}(y)-g_{n-1}(y))
  -(y^2-a_{n-1})(g_{n-1}(y)-g_{n-2}(y))\n
  &&
  -(y^2-a_{n+1})(g_{n+2}(y)-g_{n+1}(y))\,,
\end{eqnarray}
Jacobi :
\begin{eqnarray}
  X_n(x)&\!\!=\!\!&
  (\sfrac{1+\delta}{1-\delta}\,x-a_n)
  \Bigl((x-a_n)g_{n+1}(x)-(\sfrac{1+\delta}{1-\delta}\,x-a_n)g_n(x)\Bigr)\n
  &&
  -(x-a_n)
  \Bigl((x-a_{n+1})g_{n+2}(x)
  -(\sfrac{1+\delta}{1-\delta}\,x-a_{n+1})g_{n+1}(x)\Bigr)
  -(\sfrac{2\dr}{1-\delta})^2(1-x^2)g_n(x)\n
  &&
  -b_n(g_n(x)-g_{n-2}(x))+b_{n+1}(g_{n+2}(x)-g_n(x))\,,\\
  Y_n(x)&\!\!=\!\!&
  (\sfrac{1+\delta}{1-\delta}\,x-a_n)(g_{n+1}(x)-g_{n-1}(x))
  +(x-a_{n-1})g_{n-2}(x)
  -(\sfrac{1+\delta}{1-\delta}\,x-a_{n-1})g_{n-1}(x)\n
  &&
  -(x-a_{n+1})g_{n+2}(x)
  +(\sfrac{1+\delta}{1-\delta}\,x-a_{n+1})g_{n+1}(x)\,.
\end{eqnarray}
It is easy to see $X_n(x)=Y_n(x)=0$ by using the explicit forms of
$a_n,b_n$ and $g_n(x)$.

As the third step, we show the following:
\begin{prop}\label{prop:3term=feq}
The polynomial defined by the three-term recurrence
{\rm (\ref{3termrecf})} satisfies the functional equation
{\rm (\ref{feq_H})}--{\rm (\ref{feq_J})}.
\end{prop}

Therefore we obtain the following:
\begin{prop}
The polynomial defined by the functional equation
{\rm (\ref{feq_H})}--{\rm (\ref{feq_J})}
satisfies the three-term recurrence {\rm (\ref{3termrecf})}.
\end{prop}

%%%%%%%%%%%%%%%%%%%%%%%%%%%%%%%%%%%%%%%%%%%%%%%%%%%%%%%%%%%%%%%
%                                                             %
%  References                                                 %
%                                                             %
%%%%%%%%%%%%%%%%%%%%%%%%%%%%%%%%%%%%%%%%%%%%%%%%%%%%%%%%%%%%%%%

\end{document}